\documentclass[aps,prl,twocolumn,groupedaddress]{revtex4-2}

\usepackage{amsmath}
\usepackage{graphicx}

\begin{document}


\title{Fluctuation-response theorem for Kullback-Leibler divergences to quantify causation}



\author{Andrea Auconi$^{1}$}

\author{Benjamin M. Friedrich$^{1}$}

\author{Andrea Giansanti$^{2}$}

\affiliation{$1$ cfaed, Technische Universit\"at Dresden, 01069 Dresden, Germany\\$2$ Dipartimento di Fisica, Sapienza Università di Roma, 00185 Rome, Italy}



\date{\today}

\begin{abstract}
	We define a new measure of causation from a fluctuation-response theorem for Kullback-Leibler divergences, based on the information-theoretic cost of perturbations. This information response has both the invariance properties required for an information-theoretic measure and the physical interpretation of a propagation of perturbations. 
In linear systems, the information response reduces to the transfer entropy, providing a connection between Fisher and mutual information.
\end{abstract}


\maketitle

In the general framework of stochastic dynamical systems, the term \textit{causation} refers to the influence that a variable $x$ exerts over the dynamics of another variable $y$. 
Measures of causation find application in neuroscience \cite{seth2015granger}, climate studies \cite{runge2019inferring}, cancer research \cite{luzzatto2015causality}, and finance \cite{kwon2008information}. However, a widely accepted quantitative definition of causation is still missing.

Causation manifests itself in two inseparable forms: information flow \cite{ito2013information,horowitz2014thermodynamics,james2016information,auconi2017causal}, and propagation of perturbations \cite{pearl2009causality,janzing2013quantifying,aurell2016causal,baldovin2020understanding}.
Ideally, a quantitative measure of causation should connect both perspectives.

Information flow is commonly quantified by the \textit{transfer entropy} \cite{massey1990causality,schreiber2000measuring,ay2008information,parrondo2015thermodynamics,cover1999elements}, that is the average conditional mutual information corresponding to the uncertainty reduction in forecasting the time evolution of $y$ that is achieved upon knowledge of $x$. The mutual information is a special case of Kullback-Leibler (KL) divergence, a dimensionless measure of distinguishability between probability distributions \cite{amari2016information}. As such, the transfer entropy abstracts from the underlying physics to give an invariant description in terms of the strength of probabilistic dependencies.

From the interventional point of view \cite{pearl2009causality,janzing2013quantifying,aurell2016causal,baldovin2020understanding}, causation is identified with how a perturbation applied to $x$ propagates in the system to effect $y$. Although a direct perturbation of observables is unfeasible in most real-world situations, the fluctuation-response theorem establishes a connection between the response to a small perturbation and the correlation of fluctuations in the natural (unperturbed) dynamics \cite{kubo1966fluctuation,kubo1986brownian,marconi2008fluctuation,maes2020response}.

The fluctuation-response theorem considers the first-order expansion of the response with respect to the perturbation. The corresponding linear response coefficient has been suggested as a measure of causation \cite{baldovin2020understanding,aurell2016causal}. However, it has the same physical units as $y/x$, and it can assume negative values; thus, is not directly related to any information-theoretic measure.

In stochastic dynamical systems with nonlinear interactions, perturbing $x$ may not only affect the evolution of the expectation value of $y$, but it may also affect the evolution of the variance of $y$, and in fact its entire probability distribution. The KL divergence from the natural to the perturbed probability densities has recently been identified as the universal upper bound to the physical response of any observable relative to its natural fluctuations \cite{dechant2020fluctuation}. 

In this Letter, we define a new measure of causation in the form of a linear response coefficient between KL divergences, which we would like to call \textit{information response}.
In particular, we consider the ratio of two KL divergences, one for the response and one for the perturbation, where the latter represents an information-theoretic cost of the perturbation.
For small perturbations, we formulate a fluctuation-response theorem that expresses this ratio as a ratio of Fisher information.

In linear systems, this new information response reduces to the transfer entropy, which provides a connection between Fisher and mutual information, and thus a connection between fluctuation-response theory and information flows.



\paragraph{Kullback-Leibler (KL) divergence.}
Consider two probability distributions $p(w)$ and $q(w)$ of a random variable $w$. The KL divergence from $q(w)$ to $p(w)$ is defined as
\begin{eqnarray}
D\left[p(w)\big|\big|q(w)\right] \equiv \int dw~ p(w) \ln\left( \frac{p(w)}{q(w)} \right);
\end{eqnarray}
it is not symmetric in its arguments, and non-negative. Importantly, it is \textit{invariant} under invertible transformations $w\rightarrow w'$ \cite{amari2016information}, namely $D\left[p(w)\big|\big|q(w)\right]=D\left[p(w')\big|\big|q(w')\right] $.

\paragraph{The problem of causation.}
Consider a stochastic system of $n$ variables evolving with ergodic Markovian dynamics. Our goal is to \textit{define} a quantitative measure of causation, i.e., the influence that a variable $x$ exerts over the dynamics of another variable $y$. We want this definition to have both the invariance property of KL divergences, and the physical interpretation of a propagation of perturbations. 

Since the dynamics is ergodic, and therefore stationary, it suffices to consider the stochastic variables $x_0\equiv x(t=0)$, $y_0\equiv y(t=0)$ at $t=0$, and a time interval $\tau$ later $y_{\tau}\equiv y(t=\tau)$.
To avoid cluttered notation, we will implicitly assume that the current values of the remaining $n-2$ variables are absorbed into $y_0$, e.g., $p(y_{\tau}\big|y_0)\equiv p(y_{\tau}\big|y_0,z_0)$.
Conditioning on $z_0$ avoids confounding variables in $z$ to introduce spurious causal links between $x$ and $y$ \cite{runge2018causal}. 




\paragraph{Local response divergence.}
Let us consider the system at $t=0$ with steady-state distribution $p(x_0,y_0)$. We make an ideal measurement of its actual state $(x_0,y_0)$. Immediately after the measurement, we perturb the state by introducing a small displacement $\epsilon>0$ of the variable $x$, namely $x_0 \Rightarrow x_0 +\epsilon$. If the effect of this perturbation propagates to $y$, then it is reflected in the KL divergence from the natural to the perturbed prediction
\begin{multline}\label{divergence i->j}
	d^{x\rightarrow y}_{\tau}\left(x_0,y_0,\epsilon\right) \equiv  \\ D\left[p\left(y_{\tau} \big|  x_0,y_0; x_0 \Rightarrow x_0 +\epsilon \right) \big|\big|  p\left(y_{\tau} \big|  x_0,y_0 \right) \right],
\end{multline}
which is a function of the local condition $(x_0,y_0)$ and the perturbation strength $\epsilon$. We name it local response divergence, and denote its ensemble average by $\left\langle d^{x\rightarrow y}_{\tau}(x_0,y_0,\epsilon)\right\rangle$. 

The concept of causation, interpreted in the framework of fluctuation-response theory, is only meaningful with respect to an arrow of time. That means to postulate that the perturbation cannot have effects at past times
\begin{multline}\label{time arrow}
p\left(y_{\tau} \big|  x_0, y_0; x_0 \Rightarrow x_0 +\epsilon \right)\equiv\\
\begin{cases}
p\left(y_{\tau} \big|  x_0+\epsilon,y_0 \right) $  for $\tau\geq 0,\\
p\left(y_{\tau} \big|  x_0,y_0\right) $ for $\tau<0.
\end{cases}
\end{multline}
In writing the conditional probability $p\left(y_{\tau} \big|  x_0+\epsilon,y_0 \right) $, we implicitly assumed $p\left( x_0+\epsilon,y_0\right)>0$, meaning that the condition provoked by the perturbation is possible under the natural statistics. This implies that the response statistics can be predicted without actually perturbing the system, which is the main idea of fluctuation-response theory \cite{kubo1966fluctuation,kubo1986brownian,marconi2008fluctuation,maes2020response}.


\paragraph{Information-theoretic cost.}
The mean local response divergence $\left\langle d^{x\rightarrow y}_{\tau}(x_0,y_0,\epsilon)\right\rangle$, like any response function in fluctuation-response theory, is defined in relation to a perturbation, irrespective of how difficult it may be to perform this perturbation. Intuitively, we expect that it takes more effort to perturb those variables that fluctuate less.
Therefore, we consider the KL divergence from the natural to the perturbed ensemble of conditions
\begin{eqnarray}\label{divergence i}
c_x(\epsilon) \equiv D\left[p(x_0-\epsilon,y_0) \big|\big| p(x_0,y_0) \right],
\end{eqnarray}
to quantify the information-theoretic cost of perturbations, and call it \textit{perturbation divergence}. 

For example, for an underdamped Brownian particle, the perturbation divergence is equivalent to the average thermodynamic work required to perform an $\epsilon$ perturbation of its velocity, up to a factor being the temperature, see Supplementary Information (SI).
For an equilibrium ensemble in a potential $U(x)$, with Boltzmann distribution $p(x)\sim \exp(-\beta U(x))$, the perturbation divergence is the average reversible work $c_x(\epsilon) = \beta\left\langle U(x+\epsilon)-U(x) \right\rangle$.
Note that the definition of Eq. \eqref{divergence i} is general, and can be applied to more abstract models where thermodynamic quantities are not clearly identified.

\begin{center}
	\begin{figure}
		\includegraphics[trim={0.1cm 0cm 0cm 0cm},clip,scale=0.57]{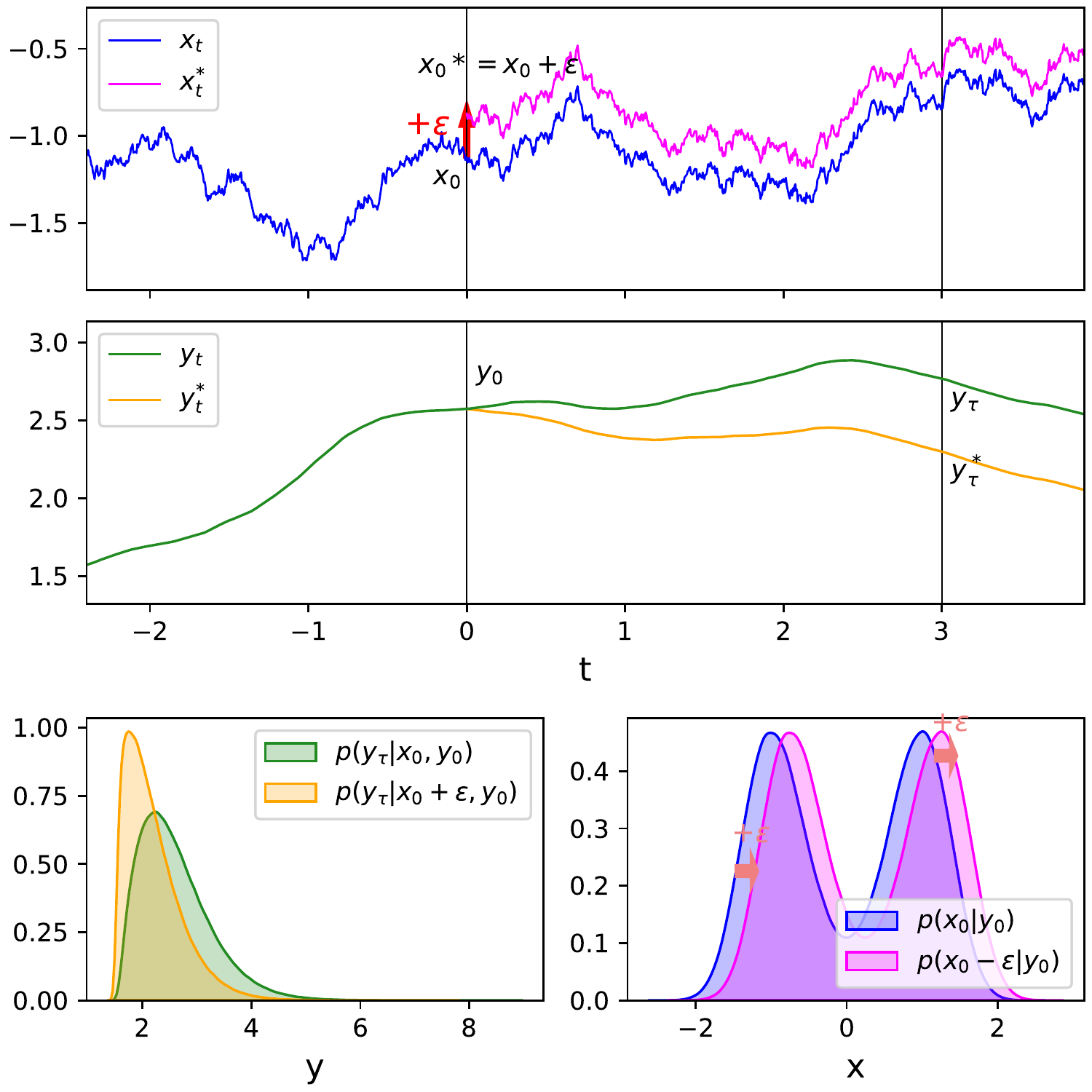}
		\caption{Here we show, on a concrete example, the origin of the two KL divergences entering the information response of Eq. \eqref{def}. (Upper) Response to the perturbation $x_0\Rightarrow x_0+\epsilon$ at the trajectory level. $x^{*}_t$ ($y^{*}_t$) is the perturbed trajectory of $x_t$ ($y_t$), for the same noise realization. (Lower Left) Local response divergence $d^{x\rightarrow y}_{\tau}(x_0,y_0,\epsilon)$: change of predicted distribution of $y_{\tau}$ for the condition $(x_0,y_0)$ for a timescale $\tau=3$. (Lower Right) Perturbation divergence $c_x(\epsilon)$: instantaneous displacement of the steady-state ensemble conditional to a particular $y_0$. The dynamics follows the nonlinear stochastic model of Eq. \eqref{nonlinear} with parameters $t_R=10$, $q=0.1$, $\alpha = 0.5$, $\beta = 0.2$, for a perturbation $\epsilon=0.25$.}
		\label{graphical_abstract}
	\end{figure}
\end{center}

\paragraph{Information response.} We introduce the information response as the ratio between mean local response divergence and perturbation divergence, in the limit of a small perturbation
\begin{eqnarray}\label{def}
	\Gamma^{x\rightarrow y}_{\tau} \equiv \lim\limits_{\epsilon\rightarrow 0} \frac{ \left\langle d^{x\rightarrow y}_{\tau}(x_0,y_0,\epsilon)\right\rangle}{c_x(\epsilon)}.
\end{eqnarray}
We can interpret $\Gamma^{x\rightarrow y}_{\tau}$ as an information-theoretic linear response coefficient.
This information response is our measure of $x\rightarrow y$ causation with respect to the timescale $\tau$, see Fig. \ref{graphical_abstract}.
The time arrow requirement (Eq. \eqref{time arrow}) implies $\Gamma^{x\rightarrow y}_{\tau}=0$ for $\tau<0$.

Introducing the \textit{local} information response $\gamma^{x\rightarrow y}_{\tau}(x_0,y_0) \equiv\lim\limits_{\epsilon\rightarrow 0} d^{x\rightarrow y}_{\tau}(x_0,y_0,\epsilon)/c_x(\epsilon)$, we can equivalently write $\Gamma^{x\rightarrow y}_{\tau} = \left\langle \gamma^{x\rightarrow y}_{\tau}(x_0,y_0) \right\rangle$. 

The information response in the form of Eq. \eqref{def} inherently relies on the concept of controlled perturbations. We can reformulate it in purely observational form, in the spirit of the fluctuation-response theorem \cite{kubo1966fluctuation,kubo1986brownian,marconi2008fluctuation,maes2020response}, provided $p(x_0,y_0,y_{\tau})$ is sufficiently smooth.

\paragraph{Fisher information.}
The one-parameter family $\{ p(y_{\tau}\big| x_0,y_0)\}_{x_0}$ of probability densities parametrized by $x_0$ (for fixed $y_0$) can be equipped with a Riemannian metric having $d^{x\rightarrow y}_{\tau}(x_0,y_0,\epsilon)$ as squared line element.  
In fact, the leading order term in the Taylor expansion of a KL divergence between probabilities that differ only by a small perturbation of a parameter is of second order, with coefficients known as Fisher information \cite{amari2016information,ito2020stochastic}.
Explicitly, expanding the mean response divergence for $\tau>0$, we obtain
\begin{multline}
	\left\langle d^{x\rightarrow y}_{\tau}(x_0,y_0,\epsilon)\right\rangle =\\ -\frac{1}{2} \epsilon^2 \left\langle \partial^2_{x_0}\ln p(y_{\tau}\big| x_0,y_0)  \right\rangle +\mathcal{O}(\epsilon^3),
\end{multline}
where we used the interventional causality requirement (Eq. \eqref{time arrow}), and probability normalization.
Similarly, for the perturbation divergence we have
\begin{eqnarray}
	c_x(\epsilon) = -\frac{1}{2}\epsilon^2 \left\langle \partial^2_{x_0}\ln p(x_0\big| y_0)  \right\rangle  +\mathcal{O}(\epsilon^3).
\end{eqnarray}

Applying the Fisher information representation to the information response, we get for $\tau>0$ 
\begin{eqnarray}\label{fd relation}
\Gamma^{x\rightarrow y}_{\tau} = \frac{\left\langle \partial^2_{x_0}\ln p\left(y_{\tau}\big| x_0,y_0\right)  \right\rangle}{\left\langle \partial^2_{x_0}\ln p\left(x_0\big| y_0\right)  \right\rangle},
\end{eqnarray}
that is the \textit{fluctuation-response theorem} for KL divergences. For generalizations and a discussion of the connection with the classical fluctuation-response theorem see \cite{Fisher} and SI text.  
Eq. \eqref{fd relation} is the ratio of two second derivatives over the same physical variable $x_0$, and it can be regarded as an application of L'H\^{o}pital's rule to Eq. \eqref{def}.

In general, Fisher information is not easily connected to Shannon entropy and mutual information \cite{wei2016mutual}. Below, we show that for linear stochastic systems, the information response, which is a ratio of Fisher information (Eq. \eqref{fd relation}), is equivalent to the transfer entropy, a conditional form of mutual information.

\paragraph{Transfer entropy.}
The most widely used measure of information flow is the conditional mutual information
\begin{multline}\label{transfer entropy}
T^{x\rightarrow y}_{\tau} \equiv \left\langle D\left[ p\left(x_0,y_{\tau}\big|y_0\right) \big|\big| p\left(x_0\big|y_0\right) p\left(y_{\tau}\big|y_0\right) \right] \right\rangle,
\end{multline}
which is generally called transfer entropy \cite{massey1990causality,schreiber2000measuring,ay2008information,parrondo2015thermodynamics,cover1999elements}.
It is the average KL divergence from conditional independence of $x_0$ and $y_{\tau}$ given $y_0$.

The transfer entropy is used in nonequilibrium thermodynamics of measurement-feedback systems, where it is related to work extraction and dissipation through fluctuation theorems \cite{sagawa2012nonequilibrium,parrondo2015thermodynamics,rosinberg2016continuous}; in data science, causal network reconstruction from time series is based on statistical significance tests for the presence of transfer entropy \cite{runge2018causal}.

If uncertainty is measured by the Shannon entropy $S[p(x)]=-\int dx ~p(x)\ln p(x)$, then the transfer entropy quantifies how much, on average, the uncertainty in predicting $y_{\tau}$ from $y_0$ decreases if we additionally get to know $x_0$, $T^{x\rightarrow y}_{\tau}=\left\langle S\left[p\left(y_{\tau}\big|y_0\right) \right]-S\left[p\left(y_{\tau}\big|x_0,y_0\right) \right]\right\rangle$.

While the joint probability $p\left(x_0,y_0,y_{\tau}\right)$ contains all the physics of the interacting dynamics of $x$ and $y$, the description in terms of the scalar transfer entropy $T^{x\rightarrow y}_{\tau}$ represents a form of coarse-graining.

We introduce the local transfer entropy $t^{x\rightarrow y}_{\tau}(x_0,y_0)= D\left[p(y_{\tau}\big|x_0,y_0)\big|\big| p(y_{\tau}\big| y_0) \right]$; thus for the (macroscopic) transfer entropy $T^{x\rightarrow y}_{\tau}=\left\langle t^{x\rightarrow y}_{\tau}(x_0,y_0) \right\rangle$.

We next show that $T^{x\rightarrow y}_{\tau}$ and $\Gamma^{x\rightarrow y}_{\tau}$ are intimately related for linear systems.

\paragraph{Linear stochastic dynamics.} As example of application, we study the information response in Ornstein-Uhlenbeck (OU) processes \cite{risken1996fokker}, i.e., linear stochastic systems of the type
\begin{eqnarray}
\frac{d\xi^{(i)}_t}{dt} + \sum_{j=1}^{n}A_{ij}\xi^{(j)}_t =\eta^{(i)}_t,
\end{eqnarray}
where $\left\langle \eta^{(i)}_t\eta^{(j)}_{t'} \right\rangle=q_{ij}\delta(t-t')$ is Gaussian white noise with symmetric and constant covariance matrix. For the system to be stationary, we require the eigenvalues of the interaction matrix $A_{ij}$ to have positive real part.
For our setting, we identify $x\equiv\xi^{(i)}$ and $y\equiv\xi^{(j)}$ for some particular $(i,j)$, and $z\equiv \{ \xi^{(k)} \}_{k=1,...,n}\backslash \{ \xi^{(i)},\xi^{(j)}\}$ as the remaining variables. 
Here, probability densities are normal distributions, $p(y_{\tau}\big| x_0,y_0) = \mathcal{N}_{y_{\tau}}(\langle y_{\tau}\big|  x_0,y_0 \rangle, \sigma^2_{y_{\tau}|  x_0,y_0})$, with mean $\langle y_{\tau}\big|  x_0,y_0 \rangle$ and variance $ \sigma^2_{y_{\tau}|  x_0,y_0} \equiv\langle y_{\tau}^2\big|  x_0,y_0 \rangle-\langle y_{\tau}\big|  x_0,y_0 \rangle^2$, and similarly for $p(y_{\tau}\big| y_0)$ and $p(x_0\big| y_0)$. Expectations depend linearly on the conditions, $\partial^2_{x_0} \langle y_{\tau}\big|  x_0,y_0 \rangle = 0$, and variances are independent of them, $\partial_{x_0}\sigma^2_{y_{\tau}|  x_0,y_0}=0$. Recall the implicit conditioning on the confounding variables $z_0$ through $y_0$.

Applying these Gaussian properties to Eq. \eqref{fd relation}, the information response becomes:
\begin{eqnarray}\label{per inf linear}
\Gamma^{x\rightarrow y}_{\tau} = \frac{\left(\partial_{x_0}\langle y_{\tau}\big|  x_0,y_0 \rangle \right)^2\sigma^2_{x_0|  y_0}}{\sigma^2_{y_{\tau}|  x_0,y_0}},
\end{eqnarray}
where $\partial_{x_0}\langle y_{\tau}\big|  x_0,y_0 \rangle$ can be interpreted as the coefficient of $x_0$ in the linear regression for $y_{\tau}$ based on the predictors $(x_0,y_0)$, and $\sigma^2_{y_{\tau}|  x_0,y_0}$ as its error variance. The variance $\sigma^2_{x_0|  y_0}$ quantifies the strength of the natural fluctuations of $x_0$ (variable to be perturbed) conditional on $y_0$ (other variables). In fact, the information-theoretic cost of the perturbation, $c_x(\epsilon)=\epsilon^2\sigma^{-2}_{x_0|  y_0}+\mathcal{O}(\epsilon^3)$, is higher if $x_0$ and $y_0$ are more correlated.

In linear systems, the transfer entropy is equivalent to Granger causality \cite{barnett2009granger} 
\begin{eqnarray}\label{TE}
	T^{x\rightarrow y}_{\tau} = \ln\left( \frac{\sigma_{y_{\tau}|  y_0}}{\sigma_{y_{\tau}|  x_0,y_0}}	 \right),
\end{eqnarray}
as can be seen by substituting the Gaussian expressions for $p(y_{\tau}\big| x_0,y_0)$ and  $p(y_{\tau}\big| y_0)$ into Eq. \eqref{transfer entropy}.

\begin{center}
\begin{figure}
\includegraphics[trim={3.5cm 8.5cm 6.0cm 9.5cm},clip,scale=0.149]{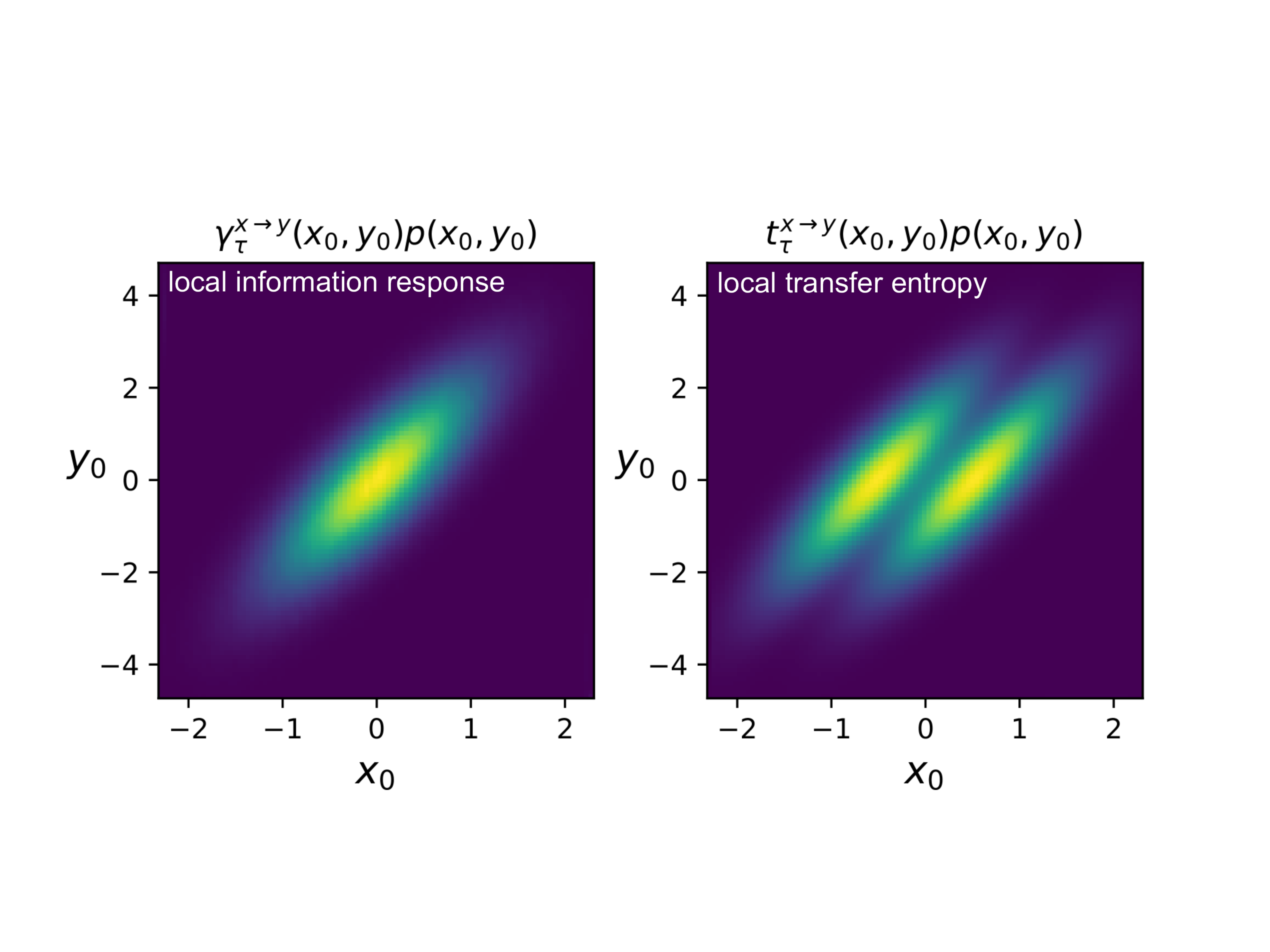}
\caption{Local information response (Left) and local transfer entropy (Right) are different, although their expectation values agree in linear systems.
The model is the OU process of Eq. \eqref{example} with parameters $t_R=10$, $q=0.1$, $\alpha=0.5$, $\beta=0.2$, observed with timescale $\tau=3$.}
\label{plot}
\end{figure}
\end{center}

The decrease in uncertainty in adding the predictor $x_0$ to the linear regression of $y_{\tau}$ based on $y_0$ reads
\begin{eqnarray}\label{linear relation}
	\sigma^2_{y_{\tau}|  y_0} -\sigma^2_{y_{\tau}|  x_0,y_0}  = \sigma^2_{x_0|  y_0} \left(\partial_{x_0}\langle y_{\tau}\big|  x_0,y_0 \rangle \right)^2,~~~
\end{eqnarray}
see SI text. Comparing Eq. \eqref{per inf linear} with Eq. \eqref{TE} and using Eq. \eqref{linear relation}, we obtain a non-trivial equivalence between information response and transfer entropy for OU processes,
\begin{eqnarray}\label{connection with TE}
	\Gamma^{x\rightarrow y}_{\tau} = e^{2T^{x\rightarrow y}_{\tau}}-1 .
\end{eqnarray}
Remarkably, despite the equivalence of the macroscopic quantities $\Gamma^{x\rightarrow y}_{\tau}$ and $T^{x\rightarrow y}_{\tau}$, the corresponding local quantities are markedly different, see Fig. \ref{plot}.

In Fig. \ref{plot}, we show the local response divergence $\gamma^{x\rightarrow y}_{\tau}(x_0,y_0)$ and local transfer entropy $t^{x\rightarrow y}_{\tau}(x_0,y_0)$ for the hierarchical OU process of two variables
\begin{eqnarray}\label{example}
	\begin{cases}
		\frac{dx}{dt} = -\frac{x}{t_R} +\eta_t,\\
		\frac{dy}{dt} = \alpha x -\beta y,
	\end{cases}
\end{eqnarray}
with $\left\langle \eta_t\eta_{t'} \right\rangle=q\delta(t-t')$, and parameters $\alpha$, $\beta>0$, $t_R>0$, $q>0$. This is possibly the simplest model of nonequilibrium stationary interacting dynamics with continuous variables  \cite{auconi2019information}. However, the pattern of Fig. \ref{plot} is qualitatively the same for any linear OU process. In fact, the perturbation $x_0\Rightarrow x_0+\epsilon$ shifts the prediction $p(y_{\tau}\big| x_0,y_0)$ by the same amount on the $y$ axis, $\epsilon\partial_{x_0}\langle y_{\tau}\big|  x_0,y_0 \rangle$, independently of the condition $(x_0,y_0)$, without affecting the variance $\sigma^2_{y_{\tau}|  x_0,y_0}$. Hence, $d^{x\rightarrow y}_{\tau}(x_0,y_0,\epsilon)$ is constant in space, and the local contribution only reflects the density $p(x_0,y_0)$, here a bivariate Gaussian.
On the contrary, the KL divergence corresponding to the change of the prediction $p(y_{\tau}\big| y_0)$ into $p(y_{\tau}\big| x_0,y_0)$ given by the knowledge of $x_0$, is strongly dependent on $(x_0,y_0)$. In fact, the local transfer entropy reads
\begin{multline}
t^{x\rightarrow y}_{\tau}(x_0,y_0)=T^{x\rightarrow y}_{\tau}+\\ \frac{\left(\partial_{x_0}\langle y_{\tau}|  x_0,y_0 \rangle \right)^2}{2\sigma^2_{y_{\tau}|  y_0}} \left[\left(x_0-\left\langle x_0\big| y_0 \right\rangle \right)^2-\sigma^2_{x_0|  y_0} \right],
\end{multline}
see SI text. In particular, for likely values $x_0\approx \left\langle x_0\big| y_0 \right\rangle$, the divergence $t^{x\rightarrow y}_{\tau}(x_0,y_0)$ is smaller compared to the unlikely situations  $x_0\gg \left\langle x_0\big| y_0 \right\rangle$ and  $x_0\ll \left\langle x_0\big| y_0 \right\rangle$. Thus, when multiplied by the steady-state density $p(x_0,y_0)$, $t^{x\rightarrow y}_{\tau}(x_0,y_0)$ attains a bimodal shape.

\paragraph{Nonlinear example.} As a counter-example for the general validity of Eq. \eqref{connection with TE} for nonlinear systems, consider the following nonlinear Langevin equation for two variables
\begin{eqnarray}\label{nonlinear}
	\begin{cases}
	\frac{dx}{dt} = -\frac{x}{t_R} +\eta_t,\\
		\frac{dy}{dt} = \alpha x^2 -\beta y.
	\end{cases}
\end{eqnarray}
Numerical simulations (same parameters as for Eq. \eqref{example}) show that Eq. \eqref{connection with TE} is violated, see SI for details. Hence, in general, the transfer entropy is not easily connected to the information response.

\paragraph{Ensemble information response.}
Similar to the above, we can define an analogous information response at the ensemble level.
From the same perturbation $x_0\Rightarrow x_0+\epsilon$, we consider the unconditional response divergence
\begin{eqnarray}\label{unconditional}
	\widetilde{d^{x\rightarrow y}_{\tau}}(\epsilon) \equiv D\left[ p\left(y_{\tau}\big|x_0\Rightarrow x_0+\epsilon\right)  \big|\big| p\left(y_{\tau}\right) \right],
\end{eqnarray}
i.e., we evaluate the response at the ensemble level, without knowledge of the measurement $(x_0,y_0)$,
\begin{eqnarray}\label{unconditional}
	p\left(y_{\tau}\big|x_0\Rightarrow x_0+\epsilon\right)
= \left\langle p\left(y_{\tau}\big|x_0,y_0;x_0\Rightarrow x_0+\epsilon\right)\right\rangle.~~~~~
\end{eqnarray}
In general $\widetilde{d^{x\rightarrow y}_{\tau}}(\epsilon)\neq \left\langle d^{x\rightarrow y}_{\tau}(x_0,y_0,\epsilon)\right\rangle$.

We define the ensemble information response as
\begin{multline}\label{ensemble causation}
\widetilde{\Gamma^{x\rightarrow y}_{\tau}} \equiv \lim\limits_{\epsilon\rightarrow 0} \frac{\widetilde{d^{x\rightarrow y}_{\tau}}(\epsilon)}{c_x(\epsilon)}\\
=-\frac{\left\langle \left\langle \partial_{x_0} \ln p\left(y_{\tau}\big|x_0,y_0\right) \big| y_{\tau} \right\rangle ^2\right\rangle}{\left\langle \partial^2_{x_0}\ln p\left(x_0\big| y_0\right)  \right\rangle},
\end{multline}
where the second line, valid only for $\tau>0$, is the corresponding fluctuation-response theorem. A straightforward generalization to arbitrary perturbation profiles $\epsilon(x_0,y_0)$ is discussed in SI text.
Note that we could write $\widetilde{d^{x\rightarrow y}_{\tau}}(\epsilon)$ through the Fisher information $\left\langle\partial^2_{\epsilon}\ln\left\langle p(y_{\tau}\big|x_0+\epsilon,y_0) \right\rangle\right\rangle\big|_{\epsilon=0}$, but the partial derivative would be over the perturbation parameter $\epsilon$, and we found it more natural to consider the self-prediction quantity $\left\langle \left\langle \partial_{x_0} \ln p\left(y_{\tau}\big|x_0,y_0\right) \big| y_{\tau} \right\rangle ^2\right\rangle$.
See SI text for technical details on expectation brakets.

In linear systems, the ensemble information response takes the form
\begin{multline}
 \widetilde{\Gamma^{x\rightarrow y}_{\tau}}~=\Gamma^{x\rightarrow y}_{\tau}e^{-2 I_{\tau}^{xy,y}}~=e^{-2 I_{\tau}^{y,y}} \left(1-e^{-2 T^{x\rightarrow y}_{\tau}}\right),
\end{multline}
where $I_{\tau}^{y,y}\equiv  D\left[p(y_0,y_{\tau})\big|\big|p(y_0)p(y_{\tau}) \right]$ is the mutual information between $y_0$ and $y_{\tau}$, and $I_{\tau}^{xy,y}= I_{\tau}^{y,y}+T^{x\rightarrow y}_{\tau}$ is the mutual information that the two predictors  $(x_0,y_0)$ together have on the output $y_{\tau}$, see SI text. 

From the nonnegativity of informations, we obtain the bound $0\leq \widetilde{\Gamma^{x\rightarrow y}_{\tau}} \leq 1$. We see that $\widetilde{\Gamma^{x\rightarrow y}_{\tau}}$ increases with the transfer entropy $T^{x\rightarrow y}_{\tau}$, and decreases with the autocorrelation $I_{\tau}^{y,y}$. 
Since $I_{\tau}^{y,y}$ diverges for $\tau\rightarrow 0$ in continuous processes, the perturbation on the $x$ ensemble takes a finite time to fully propagate its effect to the $y$ ensemble. Since time-lagged informations vanish for $\tau\rightarrow\infty$ in ergodic processes, ensembles relax asymptotically towards steady-state after a perturbation, and correspondingly the ensemble information response vanishes. This provides a trade-off shape for $\widetilde{\Gamma^{x\rightarrow y}_{\tau}}$ as a function of the timescale $\tau$.
Note the asymptotics $\widetilde{\Gamma^{x\rightarrow y}_{\tau}}/\Gamma^{x\rightarrow y}_{\tau} \rightarrow 1$  for $\tau\rightarrow \infty$, also resulting from ergodicity.

\paragraph{Discussion.} In this Letter, we introduced a new measure of causation that has both the invariance properties required for an information-theoretic measure and the physical interpretation of a propagation of perturbations. It has the form of a linear response coefficient between Kullback-Leibler divergences, and it is based on the information-theoretic cost of perturbations. We would like to call it \textit{information response}. 

We study the behavior of the information response analytically in linear stochastic systems, and show that it reduces to the known transfer entropy in this case. This establishes a first connection between fluctuation-response theory and information flow, i.e., the two main perspectives to the problem of causation at present. Additionally, it provides a new relation between Fisher and mutual information.

We suggest our information response for the design of new quantitative causal inference methods \cite{runge2018causal}.
Its practical estimation on time series, as it is normally the case for information-theoretic measures, depends on the learnability of probability distributions from a finite amount of data \cite{bialek1996field,bialek2020makes}.

\paragraph{Acknowledgments}
\begin{acknowledgments}
We thank M Scazzocchio for helpful discussions. AA is supported by the DFG through FR3429/3 to BMF;
AA, and BMF are supported through the Excellence Initiative by the German Federal and State Governments
(Cluster of Excellence PoL EXC-2068).
\end{acknowledgments}

\bibliography{Information_response}

\end{document}


\tikzstyle{every picture}+=[remember picture]


\title{Supplementary Information for the manuscript "Fluctuation-response theorem for Kullback-Leibler divergences to quantify causation"}

\author{Andrea Auconi}

\author{Benjamin Friedrich}
\author{Andrea Giansanti}



\maketitle

\begin{widetext}

\subsection{Convention on expectation symbols}
In order not to overload the formalism, when taking expectations we don't specify the variables over which they are taken. However, here we show how they can be understood immediately from the context. 
As an example, let us consider the following expression (numerator of Eq. (20) in the main text),
\begin{eqnarray}
	\left\langle \left\langle \partial_{x_0} \ln p\left(y_{\tau}\big|x_0,y_0\right) \big| y_{\tau} \right\rangle ^2\right\rangle, 
\end{eqnarray}
where $p\left(y_{\tau}\big|x_0,y_0\right)$ denotes the conditional probability of $y_{\tau}$ given the knowledge of $(x_0,y_0)$.

The term $\partial_{x_0} \ln p\left(y_{\tau}\big|x_0,y_0\right)$ is a function of the three variables $(x_0,y_0,y_{\tau})$. The expectation $\left\langle \partial_{x_0} \ln p\left(y_{\tau}\big|x_0,y_0\right) \big| y_{\tau} \right\rangle$ is conditional on the knowledge of $y_{\tau}$, therefore it is taken over the remaining variables $(x_0,y_0)$ with respect to the conditional probability $p\left(x_0,y_0\big|y_{\tau}\right)$,
\begin{eqnarray}
	\left\langle \partial_{x_0} \ln p\left(y_{\tau}\big|x_0,y_0\right) \big| y_{\tau} \right\rangle\equiv\int\int dx_0 dy_0 p\left(x_0,y_0\big|y_{\tau}\right) \partial_{x_0} \ln p\left(y_{\tau}\big|x_0,y_0\right).
\end{eqnarray}
The outer expectation, being left with only the variable $y_{\tau}$, is necessarily taken with respect to the unconditional probability $p\left(y_{\tau}\right)$,
\begin{eqnarray}
	\left\langle \left\langle \partial_{x_0} \ln p\left(y_{\tau}\big|x_0,y_0\right) \big| y_{\tau} \right\rangle ^2\right\rangle=\int dy_{\tau}  p\left(y_{\tau} \right) \left(\int\int dx_0 dy_0 p\left(x_0,y_0\big|y_{\tau}\right) \partial_{x_0} \ln p\left(y_{\tau}\big|x_0,y_0\right)\right)^2. 
\end{eqnarray}

As a second example, consider the Kullback-Leibler divergence
\begin{eqnarray}
D\left[ \left\langle p\left(y_{\tau}\big|x_0+\epsilon,y_0\right) \right\rangle  \big|\big| p(y_{\tau}) \right]\equiv \int dy_{\tau} \left\langle p\left(y_{\tau}\big|x_0+\epsilon,y_0\right) \right\rangle\ln\left(\frac{\left\langle p\left(y_{\tau}\big|x_0+\epsilon,y_0\right) \right\rangle}{p(y_{\tau})}\right),
\end{eqnarray}
which quantifies the distinguishability of two probability distributions defined over the same variable $y_{\tau}$. When introduced in its context (below Eq. \eqref{unconditional}), we know that $(x_0,y_0)$ are stochastic variables, while $\epsilon$ is a scalar. Therefore the expectation  $\left\langle p(y_{\tau}\big|x_0+\epsilon,y_0) \right\rangle$ is necessarily taken over the conditions $(x_0,y_0)$,
\begin{eqnarray}
\left\langle p\left(y_{\tau}\big|x_0+\epsilon,y_0\right) \right\rangle=\int\int dx_0 dy_0 p\left(x_0,y_0\right)p\left(y_{\tau}\big|x_0+\epsilon,y_0\right).
\end{eqnarray}

\subsubsection{Iterated conditioning theorem}
We often use the iterated conditioning theorem $\left\langle\left\langle y\big|x \right\rangle \right\rangle=\left\langle y \right\rangle$,
\begin{eqnarray}
\left\langle\left\langle y\big|x \right\rangle \right\rangle=\int dx p(x)\int dy p\left(y\big| x\right) y=\int dy p(y) y\int dx p\left(x\big|y\right)=\int dy p(y) y= \left\langle y \right\rangle.
\end{eqnarray}
For a more general proof in terms of $\sigma$-algebras, see \cite{shreve2004stochastic,karatzas1998brownian}.

Importantly, note that $\left\langle\left\langle\left\langle y\big|x \right\rangle\big|y \right\rangle\right\rangle\neq \left\langle y \right\rangle$,
\begin{eqnarray}
&\left\langle\left\langle\left\langle y\big|x \right\rangle\big|y \right\rangle\right\rangle
=\int dy p(y)\int dx p(x|y)\int dy' p(y'\big| x) y'\nonumber\\&\neq \int dy p(y)\int dx p(x)\int dy' p(y'\big| x) y'
= \int dx p(x)\int dy' p(y'\big| x) y' = \int dx p(x)\int dy p(y\big| x) y =\left\langle y \right\rangle.
\end{eqnarray}

\subsection{Perturbation divergence in an underdamped Brownian particle}
Let us consider an underdamped Brownian particle of mass $m$ immersed in a thermal reservoir at temperature $T$ and viscous damping $\lambda$. The stochastic dynamics of its velocity $v_t$ follows the Langevin equation \cite{kubo2012statistical},
\begin{equation}\label{Langevin}
	m\dot{v_t}=-\lambda v_t +\xi_t +F_t,
\end{equation}
where $\xi_t$ denotes Gaussian white noise with covariance $\left\langle \xi_t\xi_{t'} \right\rangle=2\lambda T\delta (t-t')$, and we put the Boltzmann constant to unity, $k_B=1$. The external perturbation is exerted through a force of intensity $\frac{f}{\Delta t}$ applied to the particle during the time interval $[0,\Delta t]$, that is written $F_t =\frac{f}{\Delta t} I_{[0,\Delta t]}$ and converges to a pulse for $\Delta t \rightarrow 0$.
Before the perturbation is applied at time $t=0$, the ensemble is at equilibrium and velocities are Gaussian distributed: $p(v_0)=\mathcal{G}_{v_0}(0,\frac{T}{m})$. For a generic time instant $t$ in $0\leq t\leq \Delta t$ the formal solution for $v_t$ is written
\begin{eqnarray}
&v_t = v_0 e^{-\frac{\lambda}{m}t}+\frac{1}{m}\int_{0}^{t}dt' \xi_{t'} e^{-\frac{\lambda}{m}(t-t')}+\frac{f}{m\Delta t}\int_{0}^{t}dt' e^{-\frac{\lambda}{m}(t-t')}\nonumber\\
&= v_0 e^{-\frac{\lambda}{m}t}+\frac{1}{m}\int_{0}^{t}dt' \xi_{t'} e^{-\frac{\lambda}{m}(t-t')}+\frac{f}{\lambda\Delta t}\left(1-e^{-\frac{\lambda}{m}t}\right),
\end{eqnarray}
the three terms corresponding respectively to the deterministic relaxation of the initial condition, the noise, and the perturbation.
Let us evaluate the total variation of the velocity at the end of the perturbation period,
\begin{equation}
\Delta v \equiv v_{\Delta t} -v_0 = \left(\frac{f}{\lambda\Delta t}-v_0\right)\left(1-e^{-\frac{\lambda}{m}\Delta t}\right)+\frac{1}{m}\int_{0}^{t}dt' \xi_{t'} e^{-\frac{\lambda}{m}\Delta t}.
\end{equation}
Let us consider its average over the noise realizations conditional on the initial velocity $v_0$,
\begin{equation}
\left\langle \Delta v \big| v_0  \right\rangle = \left(\frac{f}{\lambda\Delta t}-v_0\right)\left(1-e^{-\frac{\lambda}{m}\Delta t}\right) = \frac{f}{m}+\mathcal{O}(\Delta t),
\end{equation}
which is independent on $v_0$ at zero order in $\Delta t$.
The fluctuations around $\left\langle \Delta v \big| v_0  \right\rangle$ are described by the variance
\begin{eqnarray}
&\sigma^2_{\Delta v|v_0}\equiv \left\langle \left(\Delta v-\left\langle \Delta v \big| v_0  \right\rangle\right)^2 \big| v_0  \right\rangle = \frac{1}{m^2} \int_0^{\Delta t}dt\int_0^{\Delta t}dt' \left\langle \xi_{t}\xi_{t'} \right\rangle e^{-\frac{\lambda}{m}(2\Delta t -t -t')} \nonumber \\ &= \frac{T}{m}\left(1-e^{-\frac{2\lambda}{m}\Delta t}\right) = \mathcal{O}(\Delta t),
\end{eqnarray}
which vanishes in the limit of a pulse perturbation $\Delta t \rightarrow 0$. Therefore, if the perturbation is performed enough fast, the Gaussian $p\left(\Delta v \big| v_0\right)$ converges pointwise to the Dirac delta $\delta\left(\Delta v - \left\langle \Delta v \right\rangle\right)$, and we can simply write
\begin{eqnarray}
\Delta v = \left\langle \Delta v \right\rangle = \frac{f}{m}.
\end{eqnarray}

Let us now consider the amount of work required to perform the perturbation,
\begin{eqnarray}
 &W \equiv  \int_{0}^{\Delta t} F_t v_t dt  = \frac{f}{\Delta t} \int_{0}^{\Delta t} v_t dt \nonumber \\ &=  \frac{f}{\Delta t} \left[ v_0 \frac{m}{\lambda} \left(1-e^{-\frac{\lambda}{m}\Delta t}\right) +\frac{1}{m} \int_{0}^{\Delta t}dt \int_{0}^{t}dt' \xi_{t'} e^{-\frac{\lambda}{m}(t-t')} +\frac{f}{\lambda \Delta t}\left(\Delta t-\frac{m}{\lambda} \left(1-e^{-\frac{\lambda}{m}\Delta t}\right)\right) \right].
\end{eqnarray}
Its conditional expectation is
\begin{eqnarray}
 &\left\langle W \big| v_0 \right\rangle = \frac{f}{\lambda \Delta t} \left[ v_0 m \left(1-e^{-\frac{\lambda}{m}\Delta t}\right) +\frac{f}{\Delta t}\left(\Delta t-\frac{m}{\lambda} \left(1-e^{-\frac{\lambda}{m}\Delta t}\right)\right) \right]\nonumber \\ &=f\left( v_0+\frac{f}{2m} \right) +O(\Delta t )=m\Delta v \left( v_0+\frac{\Delta v}{2} \right) +\mathcal{O}(\Delta t ),
\end{eqnarray}
and its variance
\begin{eqnarray}
&\sigma^2_{W|v_0} = \left( \frac{f}{m\Delta t}\right)^2 \int_0^{\Delta t}dt\int_0^{\Delta t}ds \int_0^{t}dt' \int_0^{s}ds'  \left\langle \xi_{t'}\xi_{s'} \right\rangle e^{-\frac{\lambda}{m}(t-t'+s-s')} \nonumber \\& \leq  2\lambda T\left( \frac{f}{m\Delta t}\right)^2 \int_0^{\Delta t}dt\int_0^{\Delta t}ds \int_0^{t}dt' \leq 2\lambda T \left( \frac{f}{m}\right)^2 \Delta t = \mathcal{O}(\Delta t).
\end{eqnarray}
We see that the work performed in a pulse perturbation is $W=f\left( v_0+\frac{f}{2m} \right)$, which strongly depends on the condition $v_0$ for small $f$. The ensemble average is $\left\langle W  \right\rangle =\frac{f^2}{2m}=\frac{1}{2}m(\Delta v)^2$, and the variance $\sigma^2_W = f^2 \sigma^2_{v_0}= f^2\frac{T}{m}=2\left\langle W  \right\rangle T$. We can identify the average work $\left\langle W  \right\rangle =\frac{1}{2}m(\Delta v)^2$ with the instantaneous change in kinetic energy of the ensemble, and is therefore reversible \cite{horowitz2014second}.

Let us consider the perturbation divergence, Eq. (4) in the main text, for this 1D example,
\begin{eqnarray}\label{A1}
&c_v(\Delta v) = D\left[p(v_0-\Delta v) \big|\big| p(v_0) \right] =D\left[p(v_0) \big|\big| p(v_0+\Delta v) \right] \nonumber\\ &=   \int dv_0\mathcal{G}_{v_0}(0,\frac{T}{m}) \ln\left( \frac{\mathcal{G}_{v_0}\left(0,\frac{T}{m}\right)}{\mathcal{G}_{v_0}\left(\Delta v,\frac{T}{m}\right)} \right)  = \frac{m}{2T}  \int dv_0\mathcal{G}_{v_0}\left(0,\frac{T}{m}\right) \left(-2v_0 \Delta v +(\Delta v)^2\right) \nonumber\\ &= \frac{m}{2T}(\Delta v)^2 = \frac{\langle W \rangle}{T}.
\end{eqnarray}
For the  underdamped Brownian particle, Eq. \eqref{A1} formalizes the equivalence of information-theoretic cost and thermodynamic cost of perturbations, up to a factor being the temperature $T$.

\subsection{Proof of Equation (6) - Fisher information}
Here we review the basic connection between KL divergence and Fisher information, applied to our framework. The interventional causality requirement, Eq. (3) in the main text, imposes that for positive $\tau>0$ it holds $p\left(y_{\tau} \big|  x_0, y_0; x_0 \Rightarrow x_0 +\epsilon \right)\equiv
p\left(y_{\tau} \big|  x_0+\epsilon,y_0 \right)$, thus the local response divergence is $d^{x\rightarrow y}_{\tau}(x_0,y_0,\epsilon)=D\left[ p(y_{\tau}\big| x_0+\epsilon,y_0) \big|\big| p(y_{\tau}\big| x_0,y_0) \right]$. Let us take its ensemble average and expand in orders of the perturbation,
\begin{eqnarray}
	& \left\langle d^{x\rightarrow y}_{\tau}(x_0,y_0,\epsilon)\right\rangle = \left\langle D\left[ p(y_{\tau}\big| x_0+\epsilon,y_0) \big|\big| p(y_{\tau}\big| x_0,y_0) \right] \right\rangle \nonumber\\ &
=\int\int dx_0 dy_0 p(x_0,y_0) \int dy_{\tau}p(y_{\tau}\big| x_0+\epsilon,y_0) \ln\left(\frac{p\left(y_{\tau}\big| x_0+\epsilon,y_0\right)}{p\left(y_{\tau}\big| x_0,y_0\right)} \right) \nonumber\\ &=
\int\int dx_0 dy_0 p(x_0-\epsilon,y_0) \int dy_{\tau}p(y_{\tau}\big| x_0,y_0) \ln\left(\frac{p\left(y_{\tau}\big| x_0,y_0\right)}{p\left(y_{\tau}\big| x_0-\epsilon,y_0\right)} \right) 
\nonumber\\ &=
-\frac{1}{2} \epsilon^2 \int\int dx_0 dy_0 p(x_0-\epsilon,y_0) \int dy_{\tau}p(y_{\tau}\big| x_0,y_0) \partial^2_{x_0} \ln p\left(y_{\tau}\big| x_0,y_0\right) +\mathcal{O}(\epsilon^3)\nonumber\\ &=
-\frac{1}{2} \epsilon^2 \int\int dx_0 dy_0 p(x_0,y_0) \int dy_{\tau}p(y_{\tau}\big| x_0,y_0) \partial^2_{x_0} \ln p\left(y_{\tau}\big| x_0,y_0\right) +\mathcal{O}(\epsilon^3)
\nonumber\\ &=-\frac{1}{2} \epsilon^2 \left\langle \partial^2_{x_0}\ln p\left(y_{\tau}\big| x_0,y_0\right)  \right\rangle +\mathcal{O}(\epsilon^3),
\end{eqnarray}
where we used the probability normalization $ \int dy_{\tau}p(y_{\tau}\big| x_0,y_0) \partial_{x_0} \ln p\left(y_{\tau}\big| x_0,y_0\right)=\int dy_{\tau} \partial_{x_0}  p\left(y_{\tau}\big| x_0,y_0\right)=\partial_{x_0}\int dy_{\tau}   p\left(y_{\tau}\big| x_0,y_0\right)=\partial_{x_0}1=0$.
The term $-\left\langle \partial^2_{x_0}\ln p\left(y_{\tau}\big| x_0,y_0\right)  \right\rangle$ is called Fisher information \cite{amari2016information}.

\subsection{Proof of Equation (13)}
Here we derive the relation between the variances $\sigma^2_{y_{\tau}|x_0,y_0}$ and $\sigma^2_{y_{\tau}|y_0}$, starting from the linearity of conditional expectations, $\left\langle y_{\tau}\big| x_0,y_0 \right\rangle = x_0\partial_{x_0}\left\langle y_{\tau}\big| x_0,y_0 \right\rangle +y_0\partial_{y_0}\left\langle y_{\tau}\big| x_0,y_0 \right\rangle$, and $\left\langle x_0\big|y_0 \right\rangle=y_0\partial_{y_0}\left\langle x_0\big|y_0 \right\rangle$.
Recall that the current state of confounding variables $z_0$ is absorbed into $y_0$, so that $\left\langle y_{\tau}\big| x_0,y_0 \right\rangle\equiv\left\langle y_{\tau}\big| x_0,y_0,z_0 \right\rangle$, and $y_0\partial_{y_0}\left\langle y_{\tau}\big| x_0,y_0 \right\rangle\equiv y_0\partial_{y_0}\left\langle y_{\tau}\big| x_0,y_0,z_0 \right\rangle+z_0\partial_{z_0}\left\langle y_{\tau}\big| x_0,y_0,z_0 \right\rangle$.

We apply the iterated conditioning to $p(y_{\tau}\big| y_0)$,
\begin{eqnarray}\label{GI}
&p(y_{\tau}\big| y_0)=\int dx_0 p(x_0,y_{\tau}\big| y_0)=\int dx_0 p(x_0\big| y_0)p(y_{\tau}\big| x_0,y_0)\nonumber\\&=\int dx_0 \mathcal{G}_{x_0}\left(\left\langle x_0\big| y_0\right\rangle,\sigma^2_{x_0| y_0}\right)\mathcal{G}_{y_{\tau}}\left(\left\langle y_{\tau}\big| x_0,y_0\right\rangle,\sigma^2_{y_{\tau}| x_0,y_0}\right)\nonumber\\&=A\int dx_0 e^{-B x_0^2+C x_0} = A\sqrt{\frac{\pi}{B}}e^{\frac{C^2}{4B}},
\end{eqnarray}
where in the last line we recognized the form of a Gaussian integral with
\begin{eqnarray}
&A = \frac{1}{2\pi\sigma_{x_0| y_0}\sigma_{y_{\tau}| x_0,y_0}}\exp\left[ -\frac{\left(y_{\tau}-y_0\partial_{y_0}\left\langle y_{\tau}\big| x_0,y_0 \right\rangle \right)^2}{2\sigma^2_{y_{\tau}| x_0,y_0}}-\frac{\left\langle x_0\big|y_0 \right\rangle^2}{2\sigma^2_{x_0| y_0}} \right],\\
&B = \frac{1}{2\sigma^2_{x_0| y_0}}+\frac{\left(\partial_{x_0}\left\langle y_{\tau}\big| x_0,y_0 \right\rangle\right)^2}{2\sigma^2_{y_{\tau}| x_0,y_0}} ,\\
&C =  \frac{\left(y_{\tau}-y_0\partial_{y_0}\left\langle y_{\tau}\big| x_0,y_0 \right\rangle \right) \partial_{x_0}\left\langle y_{\tau}\big| x_0,y_0 \right\rangle}{\sigma^2_{y_{\tau}| x_0,y_0}}+\frac{\left\langle x_0\big|y_0 \right\rangle}{\sigma^2_{x_0| y_0}} .
\end{eqnarray}
Now we equate the expression of Eq. \eqref{GI} with the Gaussian $p(y_{\tau}\big| y_0)=\mathcal{G}_{y_{\tau}}\left(\left\langle y_{\tau}\big| y_0\right\rangle,\sigma^2_{y_{\tau}| y_0}\right)$, which can be done already equating the prefactors, and obtain
\begin{eqnarray}\label{E12}
\sigma^2_{y_{\tau}|y_0}-\sigma^2_{y_{\tau}|x_0,y_0}=\sigma^2_{x_0|y_0}\left( \partial_{x_0}\left\langle y_{\tau}\big| x_0,y_0 \right\rangle\right)^2,
\end{eqnarray}
which relates a reduction in variance to the corresponding linear regression coefficient.

\subsection{Proof of Equation (16)}
Here we derive the form of the local contribution $t_{\tau}^{x\rightarrow y}(x_0,y_0)$ to the transfer entropy $T_{\tau}^{x\rightarrow y}$, defined by
\begin{eqnarray}\label{def te}
T_{\tau}^{x\rightarrow y}\equiv \int dx_0dy_0dy_{\tau} p(x_0,y_0,y_{\tau})\ln \left( \frac{p(y_{\tau}|x_0,y_0)}{p(y_{\tau}|y_0)} \right)\equiv\int dx_0dy_0 p(x_0,y_0)t_{\tau}^{x\rightarrow y}(x_0,y_0),
\end{eqnarray}
where we identify $t_{\tau}^{x\rightarrow y}(x_0,y_0)=D\left[ p(y_{\tau}|x_0,y_0)\big| \big|  p(y_{\tau}|y_0) \right]$.
Substituting the Gaussian expressions $p(y_{\tau}|x_0,y_0)=\mathcal{G}(\left\langle y_{\tau}|x_0,y_0 \right\rangle, \sigma^2_{y_{\tau}|x_0,y_0})$ and $p(y_{\tau}|y_0)=\mathcal{G}(\left\langle y_{\tau}|y_0 \right\rangle, \sigma^2_{y_{\tau}|y_0})$ in $D\left[ p(y_{\tau}|x_0,y_0)\big| \big|  p(y_{\tau}|y_0) \right]$ we get
\begin{eqnarray}
&t_{\tau}^{x\rightarrow y}(x_0,y_0)=D\left[ p(y_{\tau}|x_0,y_0)\big| \big|  p(y_{\tau}|y_0) \right] \equiv \int dy_{\tau} p\left( y_{\tau}\big| x_0,y_0 \right)\ln\left(\frac{p\left( y_{\tau}\big| x_0,y_0 \right)}{p\left( y_{\tau}\big| y_0 \right)}\right)\nonumber \\
&= -\frac{1}{2}+\frac{1}{2}\ln\frac{\sigma^2_{y_{\tau}|y_0}}{\sigma^2_{y_{\tau}|x_0,y_0}}+\frac{1}{2\sigma^2_{y_{\tau}|y_0}}\int dy_{\tau} p\left( y_{\tau}\big| x_0,y_0 \right) \left( y_{\tau}-\left\langle y_{\tau}|x_0,y_0 \right\rangle+\left\langle y_{\tau}|x_0,y_0 \right\rangle-\left\langle y_{\tau}| y_0 \right\rangle \right)^2\nonumber \\
& =-\frac{1}{2}+\frac{1}{2}\ln\frac{\sigma^2_{y_{\tau}|y_0}}{\sigma^2_{y_{\tau}|x_0,y_0}}+\frac{\sigma^2_{y_{\tau}|x_0,y_0}+\left(\left\langle y_{\tau}|x_0,y_0 \right\rangle-\left\langle y_{\tau}| y_0 \right\rangle \right)^2}{2\sigma^2_{y_{\tau}|y_0}}.
\end{eqnarray}
From the linear regression $\left\langle y_{\tau}|x_0,y_0 \right\rangle=x_0\partial_{x_0} \left\langle y_{\tau}|x_0,y_0 \right\rangle+y_0\partial_{y_0} \left\langle y_{\tau}|x_0,y_0 \right\rangle$ we find
\begin{eqnarray}\label{A_avg}
&\left\langle y_{\tau}|x_0,y_0 \right\rangle-\left\langle y_{\tau}|y_0 \right\rangle=\left\langle y_{\tau}|x_0,y_0 \right\rangle- \int dx_0 p(x_0\big|y_0)\left\langle y_{\tau}|x_0,y_0 \right\rangle\nonumber \\& =\left( x_0-\left\langle x_0\big|y_0 \right\rangle\right)\partial_{x_0} \left\langle y_{\tau}|x_0,y_0 \right\rangle.
\end{eqnarray}
Transfer entropy and Granger causality are equivalent in linear systems \cite{barnett2009granger}, 
\begin{eqnarray}
T^{x\rightarrow y}_{\tau}=\frac{1}{2}\ln\left(\frac{\sigma^2_{y_{\tau}|y_0}}{\sigma^2_{y_{\tau}|x_0,y_0}}\right),
\end{eqnarray}
as can be found immediately by substituing the corresponding Gaussian expressions in Eq. \eqref{def te}.
This relation together with Eq. \eqref{E12} and Eq. \eqref{A_avg} gives
\begin{eqnarray}\label{E14}
t^{x\rightarrow y}_{\tau}(x_0,y_0)-T^{x\rightarrow y}_{\tau}
 =\frac{\left(\partial_{x_0}\langle y_{\tau}|  x_0,y_0 \rangle \right)^2}{2\sigma^2_{y_{\tau}|  y_0}} \left[\left(x_0-\left\langle x_0\big| y_0 \right\rangle \right)^2-\sigma^2_{x_0|  y_0} \right],
\end{eqnarray}
which relates the local transfer entropy $t^{x\rightarrow y}_{\tau}(x_0,y_0)$ deviation from its macroscopic counterpart $T^{x\rightarrow y}_{\tau}$ to the local conditional log-likelihood $\left(x_0-\left\langle x_0\big| y_0 \right\rangle \right)^2$.
The minimum local transfer entropy is attained at $x_0=\left\langle x_0\big| y_0 \right\rangle $ for any $y_0$, and it gives $\min_{x_0}t^{x\rightarrow y}_{\tau}(x_0,y_0)=\frac{1}{2}\left(e^{-2T^{x\rightarrow y}_{\tau}}+2T^{x\rightarrow y}_{\tau}-1 \right)\geq 0$.

Let us note that multivariate dependencies like the probabilities entering the transfer entropy, in Gaussian systems can be expressed as combinations of just bivariate correlations \cite{risken1996fokker,barrett2015exploration}.

\subsection{Proof of Eq. (20) - Ensemble information response}
Here we give a more detailed derivation of the fluctuation-response theorem for the ensemble information response. Consider the ensemble response divergence of Eq. (18-19) in the main text,
\begin{eqnarray}\label{unconditional}
&\widetilde{d^{x\rightarrow y}_{\tau}}(\epsilon) \equiv D\left[ p\left(y_{\tau}\big|x_0\Rightarrow x_0+\epsilon\right)\big|\big| p\left(y_{\tau}\right)   \right]  \nonumber\\
&=D\left[ \left\langle p\left(y_{\tau}\big|x_0,y_0;x_0\Rightarrow x_0+\epsilon\right) \right\rangle  \big|\big|p\left(y_{\tau}\right) \right]\nonumber\\
&=D\left[  \left\langle p\left(y_{\tau}\big|x_0+\epsilon,y_0\right) \right\rangle \big|\big| p\left(y_{\tau}\right)\right],
\end{eqnarray}
where the last line holds only for $\tau>0$ because of the interventional causality requirement (Eq. 3 in main text). Recall that $p(y_{\tau}\big|x_0\Rightarrow x_0+\epsilon)$ is the probability of $y_{\tau}$ given that at time $t=0$ the perturbation $x_0\Rightarrow x_0+\epsilon$ was applied, but without knowledge of the current state $(x_0,y_0)$.
Assuming $p(x_0,y_0,y_{\tau})$ to be smooth, and expanding in orders of the perturbation, from Eq. \eqref{unconditional} we obtain
\begin{eqnarray}\label{unconditional 2}
	&\widetilde{d^{x\rightarrow y}_{\tau}}(\epsilon) =\int d y_{\tau}\left\langle p\left(y_{\tau}\big|x_0+\epsilon,y_0\right) \right\rangle  \ln \left(\frac{\left\langle p\left(y_{\tau}\big|x_0+\epsilon,y_0\right) \right\rangle}{p\left(y_{\tau}\right) }\right) \nonumber\\ & = \int d y_{\tau} \left[ p\left(y_{\tau}\right)+\epsilon\left\langle \partial_{x_0} p\left(y_{\tau}\big|x_0,y_0\right)  \right\rangle \right] \ln \left( 1+\frac{\epsilon}{ p\left(y_{\tau}\right) } \left\langle \partial_{x_0} p\left(y_{\tau}\big|x_0,y_0\right)  \right\rangle +\frac{\epsilon^2}{2  p\left(y_{\tau}\right) } \left\langle \partial_{x_0}^2 p\left(y_{\tau}\big|x_0,y_0\right)  \right\rangle  \right) +\mathcal{O}(\epsilon^3) \nonumber\\ & =\frac{\epsilon^2}{2}\int dy_{\tau} \frac{1}{ p\left(y_{\tau}\right) } \left\langle \partial_{x_0} p\left(y_{\tau}\big|x_0,y_0\right)  \right\rangle ^2 +\mathcal{O}(\epsilon^3)\nonumber\\ & =\frac{\epsilon^2}{2}\left\langle \left\langle \partial_{x_0} \ln p\left(y_{\tau}\big|x_0,y_0\right) \big| y_{\tau} \right\rangle ^2\right\rangle +\mathcal{O}(\epsilon^3), 
\end{eqnarray}
where we used $\left\langle p\left(y_{\tau}\big|x_0,y_0\right) \right\rangle = p(y_{\tau})$, and from the second line we expanded the logarithm, $\ln(1+\delta)=\delta-\frac{\delta^2}{2}+\mathcal{O}(\delta^3)$, we inverted the order of integration and derivation, and used the normalization of probability, namely
\begin{eqnarray}
 \int d y_{\tau}  \left\langle \partial_{x_0}^n p(y_{\tau}\big|x_0,y_0)  \right\rangle =   \left\langle \partial_{x_0}^n \int d y_{\tau} p(y_{\tau}\big|x_0,y_0)  \right\rangle =  \left\langle \partial_{x_0}^n 1 \right\rangle = 0,
\end{eqnarray}
for any $n \in \mathbb{N}_+$. The last line of Eq. \eqref{unconditional 2} follows from
\begin{eqnarray}
 &\left\langle \partial_{x_0} p(y_{\tau}\big|x_0,y_0)  \right\rangle= \left\langle p(y_{\tau}\big|x_0,y_0)  \partial_{x_0} \ln p(y_{\tau}\big|x_0,y_0)  \right\rangle
\nonumber\\ & =\int dx_0 dy_0 p(x_0,y_0,y_{\tau})  \partial_{x_0} \ln p(y_{\tau}\big|x_0,y_0) =p(y_{\tau})\int dx_0 dy_0 p(x_0,y_0\big|y_{\tau})  \partial_{x_0} \ln p(y_{\tau}\big|x_0,y_0)
\nonumber\\ & =p(y_{\tau}) \left\langle \partial_{x_0} \ln p\left(y_{\tau}\big|x_0,y_0\right) \big| y_{\tau} \right\rangle.
\end{eqnarray}

We define the ensemble information response as
\begin{eqnarray}\label{ensemble fdt}
\widetilde{\Gamma^{x\rightarrow y}_{\tau}}(\epsilon) \equiv \lim\limits_{\epsilon\rightarrow 0} \frac{\widetilde{d^{x\rightarrow y}_{\tau}}(\epsilon)}{c_x(\epsilon)},
\end{eqnarray}
and from Eq. \eqref{unconditional 2} it directly follows the fluctuation-response theorem
\begin{eqnarray}\label{ensemble fdt}
\widetilde{\Gamma^{x\rightarrow y}_{\tau}}(\epsilon) =-\frac{\left\langle \left\langle \partial_{x_0} \ln p\left(y_{\tau}\big|x_0,y_0\right) \big| y_{\tau} \right\rangle ^2\right\rangle}{\left\langle \partial^2_{x_0}\ln p\left(x_0\big| y_0\right)  \right\rangle}.
\end{eqnarray}


\subsection{Proof of Equation (21)}

Let us substitute the Gaussian expressions for the probabilities of a linear Ornstein-Uhlenbeck process into the ensemble information response (Eq. \eqref{ensemble fdt}),
\begin{eqnarray}
&\widetilde{\Gamma^{x\rightarrow y}_{\tau}} =-\frac{\left\langle \left\langle \partial_{x_0} \ln p\left(y_{\tau}\big|x_0,y_0\right) \big| y_{\tau} \right\rangle ^2\right\rangle}{\left\langle \partial^2_{x_0}\ln p\left(x_0\big| y_0\right)  \right\rangle}\nonumber\\ &=
\sigma^2_{x_0|y_0} \left(\frac{\partial_{x_0}\left\langle y_{\tau} \big| x_0,y_0\right\rangle}{\sigma^2_{y_{\tau}|  x_0,y_0}} \right)^2 \left\langle\left\langle   y_{\tau} -\left\langle y_{\tau} \big| x_0,y_0\right\rangle\big| y_{\tau} \right\rangle^2\right\rangle \nonumber\\ &=
\frac{\Gamma^{x\rightarrow y}_{\tau} \sigma^2_{y_{\tau}}}{\sigma^2_{y_{\tau}|  x_0,y_0}} \left\langle \left(y_{\tau}- \left\langle  \left\langle y_{\tau} \big| x_0,y_0\right\rangle\big| y_{\tau} \right\rangle\right)^2\right\rangle
\nonumber\\ &=\frac{\Gamma^{x\rightarrow y}_{\tau} \sigma^2_{y_{\tau}}}{\sigma^2_{y_{\tau}|  x_0,y_0}}  \left( 1-\partial_{y_{\tau}}\langle x_0\big| y_{\tau} \rangle \partial_{x_0}\langle y_{\tau} \big| x_0,y_0\rangle  -\partial_{y_{\tau}}\langle y_0\big| y_{\tau}\rangle \partial_{y_0}\langle y_{\tau} \big| x_0,y_0\rangle  \right)^2 ,
\end{eqnarray}
where in the third passage we used $\Gamma^{x\rightarrow y}_{\tau} = \frac{\left(\partial_{x_0}\langle y_{\tau}\big|  x_0,y_0 \rangle \right)^2\sigma^2_{x_0|  y_0}}{\sigma^2_{y_{\tau}|  x_0,y_0}}$ (Eq. (11) in main text), and $\langle x_0 \big| y_{\tau} \rangle=y_{\tau}\partial_{y_{\tau}}\langle x_0\big| y_{\tau} \rangle$.
The last line can be written more compactly as
\begin{eqnarray}\label{linear itd}
\widetilde{\Gamma^{x\rightarrow y}_{\tau}}=\frac{\Gamma^{x\rightarrow y}_{\tau} \sigma^2_{y_{\tau}}}{\sigma^2_{y_{\tau}|  x_0,y_0}}  \left( 1-\partial_{y_{\tau}}\left\langle  \left\langle y_{\tau}\big| x_0,y_0\right\rangle \big| y_{\tau} \right\rangle \right)^2.
\end{eqnarray}


To relate it with Shannon information and transfer entropy, let us consider the conditional variance 
\begin{eqnarray}\label{def cond var}
\sigma^2_{y_{\tau}|  x_0,y_0}\equiv\left\langle y_{\tau}^2 \big|  x_0,y_0\right\rangle - \left\langle y_{\tau}\big| x_0,y_0\right\rangle^2.
\end{eqnarray}
Using the linear property of variances being independent of the conditions, $\partial_{x_0}\sigma^2_{y_{\tau}|  x_0,y_0}=0$, which implies $\left\langle\sigma^2_{y_{\tau}|  x_0,y_0}\right\rangle=\sigma^2_{y_{\tau}|  x_0,y_0}$, we take the expectation of Eq. \eqref{def cond var} obtaining
\begin{eqnarray}\label{squared cond var}
\sigma^2_{y_{\tau}|  x_0,y_0}\equiv\left\langle\left\langle y_{\tau}^2 \big|  x_0,y_0\right\rangle\right\rangle - \left\langle\left\langle y_{\tau}\big| x_0,y_0\right\rangle^2\right\rangle.
\end{eqnarray}
We see that the first term on the RHS is simply the unconditional variance $\sigma_{y_{\tau}}^2=\sigma_y^2$. From iterated conditioning
\begin{eqnarray}
&\left\langle\left\langle y_{\tau}^2 \big|  x_0,y_0\right\rangle\right\rangle=\int\int dx_0 dy_0 p(x_0,y_0)\left\langle y_{\tau}^2 \big|  x_0,y_0\right\rangle\nonumber\\&=\int\int\int dx_0 dy_0 dy_{\tau} p(x_0,y_0)p(y_{\tau}\big| x_0,y_0) y_{\tau}^2=\int dy_{\tau}p(y_{\tau}) y_{\tau}^2 \int\int dx_0 dy_0 p(x_0,y_0\big|y_{\tau}) \nonumber\\&=\int dy_{\tau} p(y_{\tau}) y_{\tau}^2=\left\langle y_{\tau}^2\right\rangle=\sigma_y^2.
\end{eqnarray}
Then substituting in Eq. \eqref{squared cond var} we get
\begin{eqnarray}\label{A_first}
&\sigma_y^2-\sigma^2_{y_{\tau}|  x_0,y_0}= \left\langle\left\langle y_{\tau}\big| x_0,y_0\right\rangle^2\right\rangle = \left\langle\left\langle y_{\tau}\big| x_0,y_0\right\rangle\left\langle y_{\tau}\big| x_0,y_0\right\rangle\right\rangle\nonumber\\&
=\left\langle\left\langle y_{\tau}\big| x_0,y_0\right\rangle\left(x_0\partial_{x_0}\left\langle y_{\tau}\big| x_0,y_0\right\rangle +y_0\partial_{y_0}\left\langle y_{\tau}\big| x_0,y_0\right\rangle \right)\right\rangle\nonumber\\&
= \int dx_0 dy_0 p(x_0,y_0) \left\langle y_{\tau}\big| x_0,y_0\right\rangle\left(x_0\partial_{x_0}\left\langle y_{\tau}\big| x_0,y_0\right\rangle +y_0\partial_{y_0}\left\langle y_{\tau}\big| x_0,y_0\right\rangle \right) \nonumber\\&
= \int dx_0 dy_0 p(x_0,y_0) \left(x_0\partial_{x_0}\left\langle y_{\tau}\big| x_0,y_0\right\rangle +y_0\partial_{y_0}\left\langle y_{\tau}\big| x_0,y_0\right\rangle \right)\int dy_{\tau} p\left( y_{\tau}\big| x_0,y_0\right) y_{\tau}
\nonumber\\&
= \left\langle x_0 y_{\tau}\right\rangle \partial_{x_0}\left\langle y_{\tau}\big| x_0,y_0\right\rangle +\left\langle y_0 y_{\tau}\right\rangle\partial_{y_0}\left\langle y_{\tau}\big| x_0,y_0\right\rangle
\nonumber\\&
= \sigma_y^2 \left(\partial_{y_{\tau}} \left\langle x_0\big| y_{\tau}\right\rangle \partial_{x_0}\left\langle y_{\tau}\big| x_0,y_0\right\rangle +\partial_{y_{\tau}}\left\langle y_0 \big|y_{\tau}\right\rangle\partial_{y_0}\left\langle y_{\tau}\big| x_0,y_0\right\rangle \right)
\nonumber\\&
= \sigma_y^2 \partial_{y_{\tau}}\left\langle  \left\langle y_{\tau}\big| x_0,y_0\right\rangle \big| y_{\tau} \right\rangle,
\end{eqnarray}
which substituted into Eq. \eqref{linear itd} gives
\begin{eqnarray}\label{passage}
\widetilde{\Gamma^{x\rightarrow y}_{\tau}}=\Gamma^{x\rightarrow y}_{\tau}\frac{\sigma^2_{y_{\tau}|  x_0,y_0}}{\sigma^2_{y_{\tau}}}.
\end{eqnarray}

\begin{center}
\begin{figure}
\includegraphics[trim={0.5cm 0cm 0.5cm 0.5cm},clip,scale=0.5]{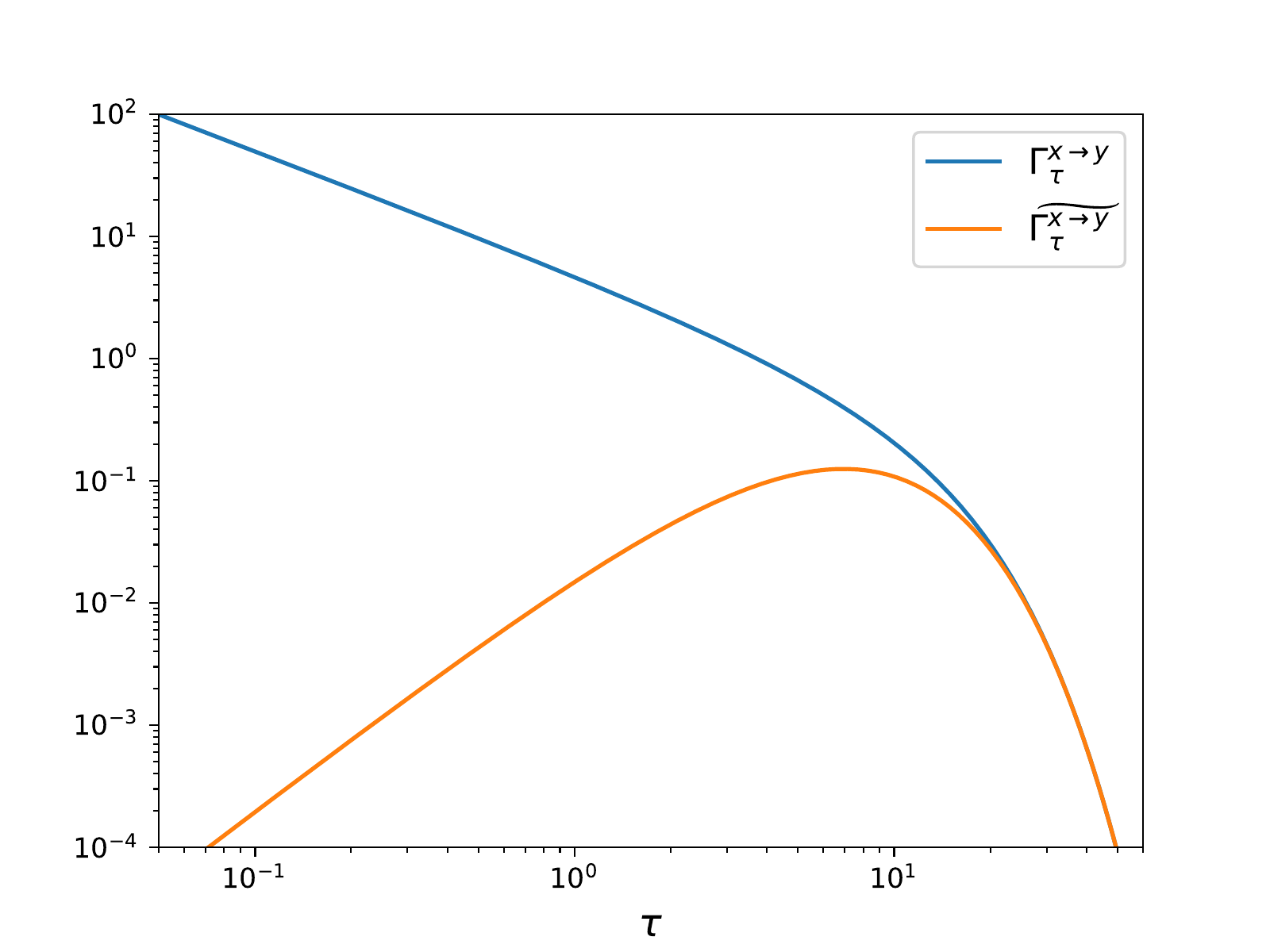}
\caption{Information response $\Gamma^{x\rightarrow y}_{\tau}$ and its ensemble counterpart $\widetilde{\Gamma^{x\rightarrow y}_{\tau}}$ as a function of the timescale $\tau$. The model is the (linear) OU process of Eq. \eqref{example} with parameters $t_R=10$, $\beta=0.2$, $\alpha=0.5$, $D=0.1$.}
\label{FIG A1}
\end{figure}
\end{center}

Let us introduce the total information as the mutual information between the couple of variables $(x_0,y_0)$ and $y_{\tau}$,
\begin{eqnarray}
 &I_{\tau}^{xy,y}\equiv D\left[ p\left(x_0,y_0,y_{\tau}\right)\big|\big| p\left(x_0,y_0\right)p\left(y_{\tau}\right) \right]\nonumber\\
&=D\left[ p\left(y_0,y_{\tau}\right)\big|\big| p\left(y_0\right)p\left(y_{\tau}\right) \right]+\left\langle D\left[ p\left(x_0,y_{\tau}\big|y_0\right)\big|\big| p\left(x_0\big|y_0\right)p\left(y_{\tau}\big|y_0\right) \right]\right\rangle\nonumber\\
&=I_{\tau}^{y,y}+T^{x\rightarrow y}_{\tau}.
\end{eqnarray}

In linear systems the mutual information is $I_{\tau}^{y,y}=\frac{1}{2}\ln\left(\frac{\sigma^2_{y_{\tau}}}{\sigma^2_{y_{\tau}| y_0}}\right)$ and the transfer entropy $T^{x\rightarrow y}_{\tau}  =\frac{1}{2}\ln\left(\frac{\sigma^2_{y_{\tau}|  y_0}}{\sigma^2_{y_{\tau}|  x_0,y_0}}\right)$, so that the total information is $I_{\tau}^{xy,y}=I_{\tau}^{y,y}+T^{x\rightarrow y}_{\tau}=\frac{1}{2}\ln\left(\frac{\sigma^2_{y_{\tau}}}{\sigma^2_{y_{\tau}|  x_0,y_0}}\right)$, and from Eq. \eqref{passage} we obtain
\begin{eqnarray}
 \widetilde{\Gamma^{x\rightarrow y}_{\tau}}~=\Gamma^{x\rightarrow y}_{\tau}e^{-2 I_{\tau}^{xy,y}}~=e^{-2 I_{\tau}^{y,y}} \left(1-e^{-2 T^{x\rightarrow y}_{\tau}}\right),
\end{eqnarray}
which relates the two definitions of information response.
In Fig. \ref{FIG A1} we plot them for the 2D hierarchical OU process \cite{auconi2019information} (Eq. 15 in the main text)
\begin{eqnarray}\label{example}
	\begin{cases}
		\frac{dx}{dt} = -\frac{x}{t_R} +\eta_t,\\
		\frac{dy}{dt} = \alpha x -\beta y,
	\end{cases}
\end{eqnarray}
with $\left\langle \eta_t\eta_{t'} \right\rangle=q\delta(t-t')$, and parameters $\alpha$, $\beta>0$, $t_R>0$, $q>0$.

\subsection{Nonlinear example}
We considered the nonlinear SDE
\begin{eqnarray}\label{nonlinear}
	\begin{cases}
		\frac{dx}{dt} = -\frac{x}{t_{rel}} +\eta_t,\\
		\frac{dy}{dt} = \alpha x^2 -\beta y,
	\end{cases}
\end{eqnarray}
with white noise $\left\langle \eta_t\eta_{t'} \right\rangle=q\delta(t-t')$, and parameters $\alpha$, $\beta>0$, $t_{rel}>0$, $q>0$.
For intuition, $x$ can be interpreted as an external fluctuating concentration signal with timescale $t_{rel}$, and $y$ as a noiseless biochemical response that is more activated when the signal is far from its average value $x=0$ in either positive or negative direction. We checked numerically that the equivalence between transfer entropy and information response for linear OU processes (Eq. 14 in the main text) does not hold here (see Fig. \ref{plot_tau_nonlinear}), and the transfer entropy is not easily connected to interventional causation.
For a specific $\tau=3$ we plot the local contributions to the response divergence and transfer entropy, see Fig. \ref{plot}. The local response divergence is governed, at least qualitatively, by the squared derivative of the quadratic interaction $\sim\left(\partial_x  x^2 \right)^2\sim x^2$. As a result the product $d^{x\rightarrow y}_{\tau}(x_0,y_0,\epsilon)p(x_0,y_0)$ is bimodal. The conditional local density $p(x_0\big| y_0)$, at least for large $y_0$, is also bimodal because of the quadratic driving and finite correlation time of the signal. For a given $y_0$, the local transfer entropy $ t^{x\rightarrow y}_{\tau}(x_0,y_0)\equiv D\left[p(y_{\tau}\big|x_0,y_0)\big|\big| p(y_{\tau}\big| y_0) \right]  $ is larger for unlikely $x_0$ which means, given the bimodality of $p(x_0\big| y_0)$, in addition to the increase in the two sides $x_0\rightarrow \pm \infty$ like it is also in the linear case, also towards a peak at $x=0$. Therefore, when multiplied by the local density  $p(x_0, y_0)$, the local transfer entropy contribution $t^{x\rightarrow y}_{\tau}(x_0,y_0)p(x_0,y_0)$ has four peaks for a fixed $y_0$ (three for small $y_0$).

\begin{center}
\begin{figure}
\includegraphics[trim={3.5cm 8.5cm 6.0cm 9.5cm},clip,scale=0.25]{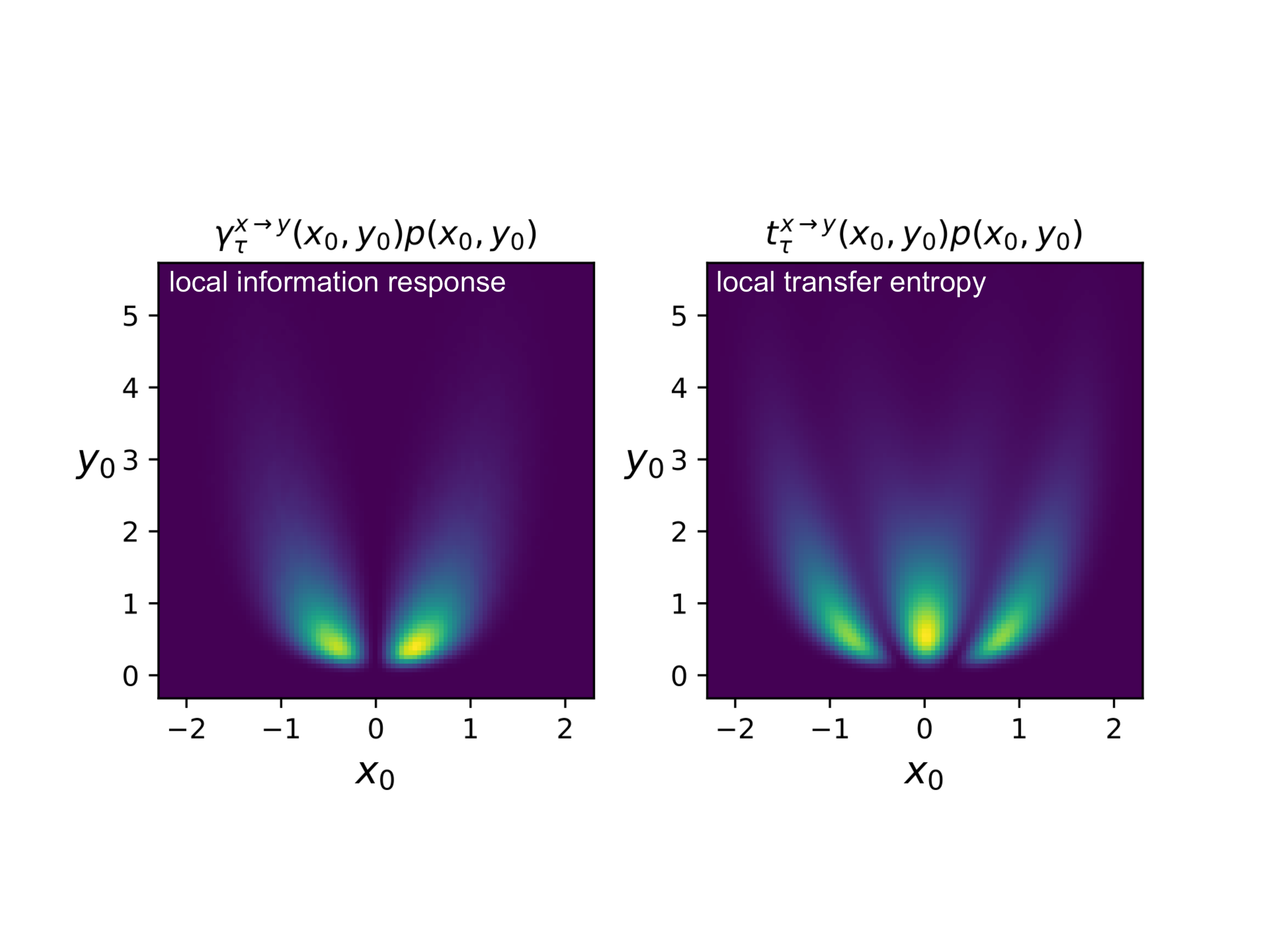}
\caption{Local information response and local transfer entropy in the nonlinear model of Eq. \eqref{nonlinear} with parameters $t_{rel}=10$, $\beta=0.2$, $\alpha=0.5$, $D=0.1$, for a timescale $\tau=3$.}
\label{plot}
\end{figure}
\end{center}

\begin{center}
\begin{figure}
\includegraphics[trim={0.5cm 0cm 0.5cm 0.5cm},clip,scale=0.78]{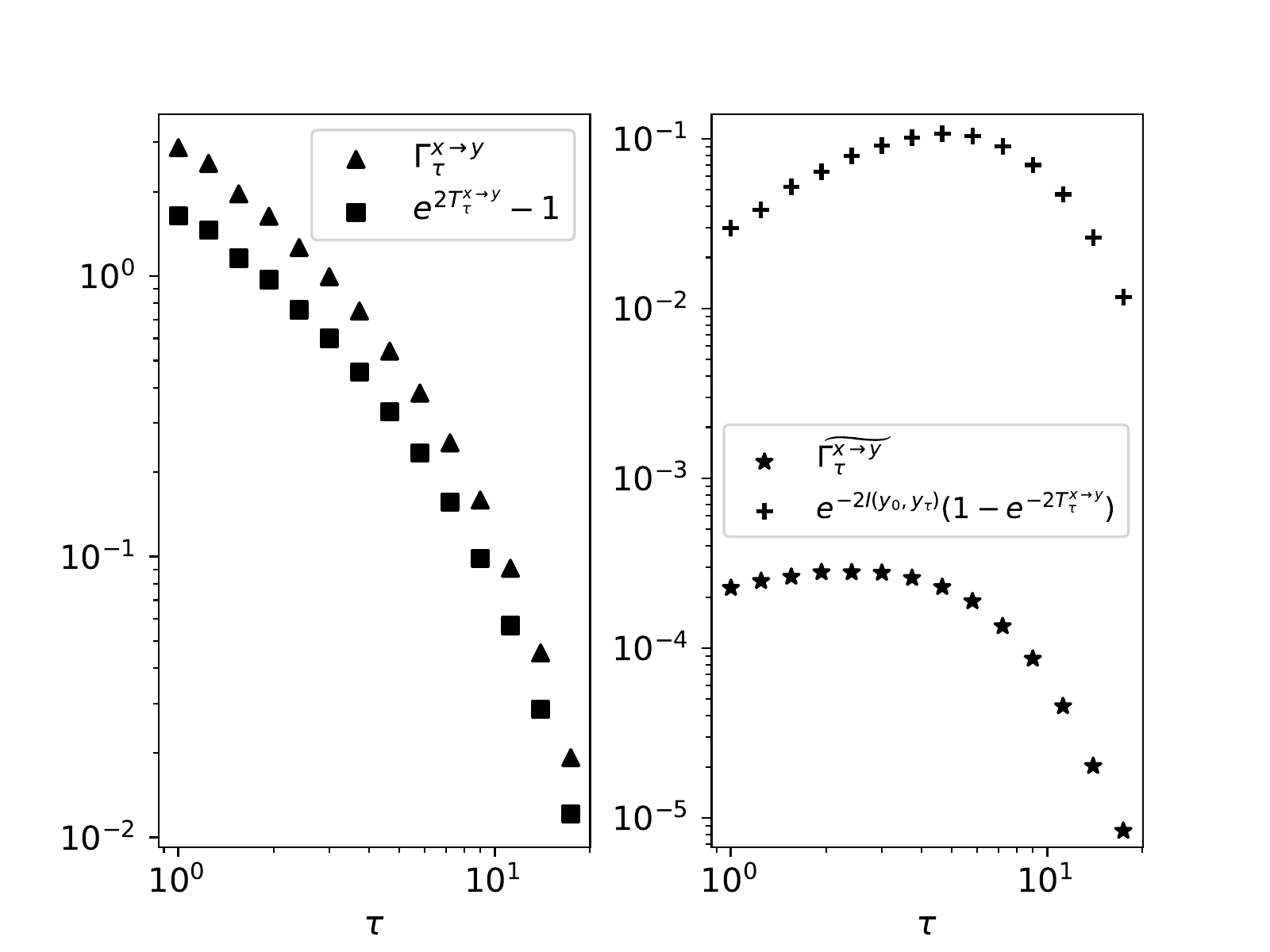}
\caption{Information response $\Gamma^{x\rightarrow y}_{\tau}$ and its ensemble counterpart $\widetilde{\Gamma^{x\rightarrow y}_{\tau}}$ as a function of the timescale $\tau$, compared to the corresponding combination of mutual informations they reduce to in linear systems. The model is the nonlinear Langevin system of Eq. \eqref{nonlinear} with parameters $t_{rel}=10$, $\beta=0.2$, $\alpha=0.5$, $D=0.1$.}
\label{plot_tau_nonlinear}
\end{figure}
\end{center}

\subsection{General perturbations}
This manuscript is based on a particular type of perturbation, namely an $\epsilon$-shift of a variable at $t=0$. In the local response divergence, since the measurement completely resolves the uncertainty, the perturbation corresponds to a shift of the corresponding delta distribution, $\delta \left(x(0)-x_0\right)\delta \left(y(0)-y_0\right)\Rightarrow \delta \left(x(0)-x_0-\epsilon\right)\delta \left(y(0)-y_0\right)$. In the ensemble response divergence instead the perturbation is written $p\left(x_0,y_0\right)\Rightarrow p\left(x_0-\epsilon,y_0\right)$. Note that in both cases we use the information-theoretic cost at the ensemble level, $c_x\equiv D\left[ p\left(x_0,y_0\right)\big|\big| p\left(x_0-\epsilon,y_0\right) \right]$, since the KL divergence between two different Dirac-deltas is not defined.

More in general, a perturbation of $x_0$ at the ensemble level can be written in the form
\begin{eqnarray}
p\left(x_0\big|y_0\right)\Rightarrow p\left(x_0\big|y_0\right) \left[1+\epsilon h_x\left(x_0,y_0\right)\right] \equiv p^*\left(x_0\big|y_0\right),
\end{eqnarray}
with $\int dx_0 p\left(x_0\big|y_0\right) h_x\left(x_0,y_0\right)=0$.
The perturbed probability of $y_{\tau}$ is written
\begin{eqnarray}
&p^*\left(y_{\tau}\right)= \int\int dx_0 dy_0  p\left(y_{\tau}\big| x_0,y_0\right)p\left(y_{0}\right)p^*\left(x_0\big|y_0\right)\nonumber\\&= \int\int dx_0 dy_0  p\left(y_{\tau}\big| x_0,y_0\right)p\left(y_{0}\right)p\left(x_0\big|y_0\right) \left[1+\epsilon h_x\left(x_0,y_0\right)\right]\nonumber\\&= p\left(y_{\tau}\right)\left[1+\epsilon \int\int dx_0 dy_0  p\left(x_0,y_0\big| y_{\tau}\right) h_x\left(x_0,y_0\right)\right]\nonumber\\&= p\left(y_{\tau}\right)\left[1+\epsilon \left\langle h_x\left(x_0,y_0\right) \big| y_{\tau}\right\rangle \right].
\end{eqnarray}
We can define the generalized response divergence as a functional
\begin{eqnarray}
&d_{\tau}^{i\rightarrow j}[h](\epsilon) \equiv D\left[ p^*\left(y_{\tau}\right)  \big|\big| p\left(y_{\tau}\right) \right]\nonumber\\&= \int d y_{\tau} p(y_{\tau})\left[1+\epsilon \left\langle h_x\left(x_0,y_0\right) \big| y_{\tau}\right\rangle \right] \ln\left(1+\epsilon \left\langle h_x\left(x_0,y_0\right) \big| y_{\tau}\right\rangle \right) \nonumber\\&= \int d y_{\tau} p(y_{\tau}) \left(\epsilon \left\langle h_x\left(x_0,y_0\right) \big| y_{\tau}\right\rangle+\frac{\epsilon^2}{2} \left\langle h_x\left(x_0,y_0\right) \big| y_{\tau}\right\rangle^2 \right) +\mathcal{O}(\epsilon^3)\nonumber\\&=
\epsilon \left\langle \left\langle h_x\left(x_0,y_0\right) \big| y_{\tau}\right\rangle\right\rangle+\frac{\epsilon^2}{2} \left\langle\left\langle h_x\left(x_0,y_0\right) \big| y_{\tau}\right\rangle^2\right\rangle\nonumber\\&=
\frac{\epsilon^2}{2} \left\langle\left\langle h_x\left(x_0,y_0\right) \big| y_{\tau}\right\rangle^2\right\rangle,
\end{eqnarray}
where in the last passage we used the iterated conditioning theorem and the $h(x_0,y_0)$ normalization, $ \left\langle \left\langle h_x\left(x_0,y_0\right) \big| y_{\tau}\right\rangle\right\rangle=  \left\langle h_x\left(x_0,y_0\right) \right\rangle= \left\langle\int dx_0 p\left(x_0\big|y_0\right) h_x\left(x_0,y_0\right)\right\rangle=0$. 
Similarly, the information-theoretic cost is
\begin{eqnarray}
&c_x[h](\epsilon) \equiv D\left[ p^*\left(x_0\big|y_0\right)p(y_0) \big|\big| p\left(x_0,y_0\right) \right]
= \left\langle D\left[ p^*\left(x_0\big|y_0\right) \big|\big| p\left(x_0\big| y_0\right) \right] \right\rangle \nonumber\\&=
 \left\langle \int dx_0 p\left(x_0\big| y_0\right)\left[1+\epsilon h_x\left(x_0,y_0\right)\right] \ln\left[1+\epsilon h_x\left(x_0,y_0\right)\right] \right\rangle \nonumber\\&= \frac{\epsilon^2}{2} \left\langle \int dx_0 p\left(x_0\big| y_0\right)  h_x^2\left(x_0,y_0\right) \right\rangle = \frac{\epsilon^2}{2} \left\langle  h_x^2\left(x_0,y_0\right) \right\rangle.
\end{eqnarray}
Then the generalized information response and its corresponding fluctuation-response theorem are written
\begin{eqnarray}
\widetilde{\Gamma^{x\rightarrow y}_{\tau}}[h] \equiv \lim\limits_{\epsilon\rightarrow 0} \frac{d_{\tau}^{i\rightarrow j}[h](\epsilon)}{c_x[h](\epsilon)}=\frac{\left\langle\left\langle h_x\left(x_0,y_0\right) \big| y_{\tau}\right\rangle^2\right\rangle}{\left\langle  h_x^2\left(x_0,y_0\right) \right\rangle}.
\end{eqnarray}

\subsection{Linear response theory}

In this section we review the classical fluctuation-response theorem \cite{risken1996fokker,kubo1966fluctuation,kubo1986brownian,marconi2008fluctuation,maes2020response} and the linear fluctuation-response inequality for the corresponding KL divergence \cite{dechant2020fluctuation}, and we motivate the introduction of the information response in this framework. Let us expand the average response of $y_{\tau}$ to the small perturbation $x_0\Rightarrow x_0+\epsilon$, for $\tau>0$,
\begin{eqnarray}
&\left\langle y_{\tau} \big| x_0\Rightarrow x_0+\epsilon \right\rangle = \left\langle \left\langle y_{\tau} \big| x_0+\epsilon,y_0 \right\rangle\right\rangle\equiv\int\int \int dx_0 dy_0 dy_{\tau}~y_{\tau} p(x_0,y_0)p(y_{\tau}\big|x_0+\epsilon,y_0)  \nonumber\\&=\int\int \int dx_0 dy_0 dy_{\tau} ~y_{\tau} p(y_{\tau}\big|x_0,y_0) p(x_0-\epsilon,y_0) \nonumber\\&= \int\int \int dx_0 dy_0 dy_{\tau}~ y_{\tau} p(y_{\tau}\big|x_0,y_0) \left[ p(x_0,y_0) -\epsilon\partial_{x_0}p(x_0,y_0) +\mathcal{O}(\epsilon^2) \right]\nonumber \\& =\left\langle y_{\tau} \right\rangle -\epsilon \left\langle y_{\tau} \partial_{x_0}\ln p(x_0,y_0) \right\rangle +\mathcal{O}(\epsilon^2).
\end{eqnarray}
In the limit $\epsilon\rightarrow 0$ we obtain the \textit{fluctuation-response theorem}:
\begin{eqnarray}\label{fluctuation-response theorem}
\lim_{\epsilon\rightarrow 0} \frac{\left\langle y_{\tau} \big| x_0\Rightarrow x_0+\epsilon \right\rangle-\left\langle y_{\tau}\right\rangle}{\epsilon} = -\left\langle y_{\tau} \partial_{x_0}\ln p(x_0,y_0) \right\rangle,
\end{eqnarray}
which equates the linear response coefficient to a correlation evaluated in the unperturbed dynamics.

For those systems having a symmetry in the correlation function, $\left\langle y_{\tau} \partial_{x_0}\ln p(x_0,y_0) \right\rangle=\pm \left\langle y_{-\tau} \partial_{x_0}\ln p(x_0,y_0) \right\rangle$, the Wiener-Kintchine theorem applied to Eq. \eqref{fluctuation-response theorem} gives the equivalence between subseptibility and cross-spectral density, that applied to Brownian motion gives the celebrated Einstein relation \cite{risken1996fokker}.

Let us now take the absolute value of both sides in the fluctuation-response theorem (Eq. \eqref{fluctuation-response theorem}), apply the iterated conditioning to the RHS, and then the Cauchy-Schwarz inequality $\big| \int f(x) g(x) dx \big|^2 \leq  \int \big| f(x) \big|^2 dx\int \big| g(x) \big|^2 dx$, to obtain
\begin{eqnarray}
&\bigg| \lim_{\epsilon\rightarrow 0} \frac{\left\langle y_{\tau} \big| x_0\Rightarrow x_0+\epsilon \right\rangle-\left\langle y_{\tau}\right\rangle}{\epsilon}\bigg| = \bigg| \left\langle y_{\tau} \partial_{x_0}\ln p(x_0,y_0) \right\rangle\bigg|= \bigg| \left\langle y_{\tau}  \left\langle \partial_{x_0}\ln p(x_0,y_0) \big| y_{\tau}\right\rangle \right\rangle\bigg|\nonumber\\ &= \bigg| \left\langle \left( y_{\tau}-\left\langle y_{\tau} \right\rangle\right) \left\langle \partial_{x_0}\ln p(x_0,y_0) \big| y_{\tau}\right\rangle \right\rangle\bigg|  =  \bigg| \int dy_{\tau} p(y_{\tau})\left( y_{\tau}-\left\langle y_{\tau} \right\rangle\right)  \left\langle \partial_{x_0}\ln p(x_0,y_0) \big| y_{\tau}\right\rangle \bigg| \nonumber\\ &\leq \sqrt{ \int dy_{\tau} p(y_{\tau}) \left( y_{\tau}-\left\langle y_{\tau} \right\rangle\right)^2 \int dy_{\tau} p(y_{\tau}) \left\langle \partial_{x_0}\ln p(x_0,y_0) \big| y_{\tau}\right\rangle ^2 } \nonumber\\ &= \sqrt{ \sigma_{y_{\tau}}^2 \left\langle\left\langle \partial_{x_0}\ln p(x_0,y_0) \big| y_{\tau}\right\rangle ^2\right\rangle },
\end{eqnarray}
where we used $\left\langle \left\langle y_{\tau} \right\rangle \left\langle \partial_{x_0}\ln p(x_0,y_0) \big| y_{\tau}\right\rangle \right\rangle=\left\langle y_{\tau} \right\rangle \left\langle \left\langle \partial_{x_0}\ln p(x_0,y_0) \big| y_{\tau}\right\rangle \right\rangle=\left\langle y_{\tau} \right\rangle \left\langle \partial_{x_0}\ln p(x_0,y_0)  \right\rangle=0$, and identified the variance $\sigma_{y_{\tau}}^2=\int dy_{\tau} p(y_{\tau}) \left( y_{\tau}-\left\langle y_{\tau} \right\rangle\right)^2$. Using the expressions for the ensemble response divergence of Eq. \eqref{unconditional}-\eqref{unconditional 2}, namely $\widetilde{d^{x\rightarrow y}_{\tau}}(\epsilon) \equiv D\left[  p(y_{\tau}\big|x_0\Rightarrow x_0+\epsilon) \big|\big| p(y_{\tau})  \right]=\frac{\epsilon^2}{2}\left\langle\left\langle \partial_{x_0}\ln p(x_0,y_0) \big| y_{\tau}\right\rangle ^2\right\rangle +\mathcal{O}(\epsilon^3)$, we obtain the \textit{linear fluctuation-response inequality} \cite{dechant2020fluctuation}
\begin{eqnarray}
\big| \left\langle y_{\tau} \big| x_0\Rightarrow x_0+\epsilon \right\rangle-\left\langle y_{\tau}\right\rangle \big| \leq \sigma_{y_{\tau}}\sqrt{  2D\left[  p(y_{\tau}\big|x_0\Rightarrow x_0+\epsilon) \big|\big| p(y_{\tau})  \right] }+\mathcal{O}(\epsilon^\frac{3}{2}),
\end{eqnarray}
which identifies the KL divergence $D\left[  p(y_{\tau}\big|x_0\Rightarrow x_0+\epsilon) \big|\big| p(y_{\tau})  \right]$ as the information-theoretic bound to the response of $y_{\tau}$ relative to its natural fluctuations $\sigma_{y_{\tau}}$.

The two fundamental results derived above suggest the possibility of a fluctuation-response theorem for KL divergences, that is what we do in the main text.  In particular, starting from the KL divergence $D\left[  p(y_{\tau}\big|x_0\Rightarrow x_0+\epsilon) \big|\big| p(y_{\tau})  \right]$ which describes the response, we define a second KL divergence to quantify the information-theoretic cost of perturbations. Then we expand separately these two KL divergences, and they are both of order $\mathcal{O}(\epsilon^2)$ for $\epsilon\rightarrow 0$, with the corresponding Taylor coefficients having the form of Fisher information. The resulting linear response coefficient is then a ratio of Fisher information, such relation we interpret as an information-theoretic fluctuation-response theorem.

Here we sketch the analogy between ours and the classical fluctuation-response theorem:
\begin{equation*}
\lim_{\epsilon\rightarrow 0}\frac{\tikz[baseline]{
            \node[fill=blue!20,anchor=base] (t1)
            {$ \left\langle y_{\tau} \big| x_0\rightarrow x_0+\epsilon \right\rangle-\left\langle y_{\tau}\right\rangle$};
        } }{\tikz[baseline]{
            \node[fill=red!20, ellipse,anchor=base] (t2)
            {$\epsilon$};
        } } =  \tikz[baseline]{
            \node[fill=green!20,anchor=base] (t3)
            {$-\left\langle y_{\tau} \partial_{x_0}\ln p(x_0,y_0) \right\rangle$};
        }.    
\end{equation*}

\begin{itemize}
     \item \contour{blue!20}{Response}
        \tikz \node[coordinate] (n1) {};
    \item \contour{red!20}{Perturbation}
        \tikz\node [coordinate] (n2) {};
    \item \contour{green!20}{Correlation}
        \tikz\node [coordinate] (n3) {};
\end{itemize}

\begin{equation*}
\widetilde{\Gamma_{\tau}^{x\rightarrow y}}\equiv\lim_{\epsilon\rightarrow 0}\frac{\tikz[baseline]{
            \node[fill=blue!20,anchor=base] (t4)
            {$ D\left[ \left\langle p\left(y_{\tau}\big|x_0+\epsilon,y_0\right) \right\rangle  \big|\big| p(y_{\tau})\right]$};
        } }{\tikz[baseline]{
            \node[fill=red!20, ellipse,anchor=base] (t5)
            {$D\left[ p(x_0-\epsilon,y_0) \big|\big| p(x_0,y_0) \right]$};
        } } =  \tikz[baseline]{
            \node[fill=green!20,anchor=base] (t6)
            {$- \frac{\left\langle \left\langle \partial_{x_0} \ln p\left(y_{\tau}\big|x_0,y_0\right) \big| y_{\tau} \right\rangle ^2\right\rangle}{\left\langle \partial^2_{x_0}\ln p\left(x_0\big| y_0\right)  \right\rangle}$};}\stackrel{linear}{=}
 \tikz[baseline]{
            \node[fill=green!20,anchor=base]
            {$e^{-2 I_{\tau}^{y,y}} (1-e^{-2 T^{x\rightarrow y}_{\tau}})$};
        },
\begin{tikzpicture}[overlay]
        \path[->] (n1) edge [out=45,in=170] (t1);
        \path[->] (n2) edge  (t2);
        \path[->] (n3) edge [bend right] (t3);
        \path[->] (n1) edge  (t4);
        \path[->] (n2) edge [out=-60,in=170] (t5);
        \path[->] (n3) edge [bend left] (t6);
\end{tikzpicture}
   \end{equation*}
and we added the connection between fluctuation-response theory and mutual informations obtained for linear systems (Eq. (21) in the main text).

We outlined the analogy of the classical fluctuation-dissipation theorem with our ensemble information response $\widetilde{\Gamma_{\tau}^{x\rightarrow y}}$, but in the main text we first focus on the information response $\Gamma_{\tau}^{x\rightarrow y}$, which is the averaged conditional (local) version of $\widetilde{\Gamma_{\tau}^{x\rightarrow y}}$. While the connection with the original fluctuation-response theorem is loose, the structure of perturbation-response-correlation is analogous,
\begin{equation*}
\Gamma_{\tau}^{x\rightarrow y}\equiv\lim_{\epsilon\rightarrow 0}\frac{\tikz[baseline]{
            \node[fill=blue!20,anchor=base] 
            {$ \left\langle D\left[  p\left(y_{\tau}\big|x_0+\epsilon,y_0\right) \big|\big| p\left(y_{\tau}\big|x_0,y_0\right) \right]\right\rangle$};
        } }{\tikz[baseline]{
            \node[fill=red!20, ellipse,anchor=base] 
            {$D\left[p(x_0,y_0) \big|\big| p(x_0-\epsilon,y_0) \right]$};
        } } =  \tikz[baseline]{
            \node[fill=green!20,anchor=base] 
            {$\frac{\left\langle \partial^2_{x_0} \ln p\left(y_{\tau}\big|x_0,y_0\right) \right\rangle}{\left\langle \partial^2_{x_0}\ln p\left(x_0\big| y_0\right)  \right\rangle}$};
        }\stackrel{linear}{=}
 \tikz[baseline]{
            \node[fill=green!20,anchor=base]
            {$e^{2T_{\tau}^{x\rightarrow y}}-1$};
        }.
\end{equation*}

\subsection{Application to data science}

In the main text, we present our results in relation to the current literature in theoretical fields such as fluctuation-response theory, information theory, and nonequilibrium thermodynamics. Here we motivate our study also in relation to the current trend of data science.

The accuracy of predictions is one of the main goals in statistics and applied physics.
In general, predictions can be obtained from mechanical models, where physical intuition plays a role in selecting the relevant observables and characterizing their interactions  \cite{vulpiani2020effective}, or from machine learning approaches, where the large availability of (labeled) data enables high-dimensional computing architectures to be trained for pattern recognition  \cite{lecun2015deep}.
In this latter case, predictions are not explainable in terms of intuitive mechanisms or geometrical relations. In other words, the ability of doing predictions does not imply understanding \cite{marcus2020gpt2,goebel2018explainable}. 

With the aim of \textit{explainability}, an helpful representation of the dynamics is given by \textit{causal networks} \cite{ito2013information,auconi2019information}, where weighted directed links between nodes represent the propagation of perturbations between variables in the network, or the information flow. Causal networks are coarse-grained representations of the dynamics and its interactions, limited to a set of scalars representing how much a variable is influencing the dynamics of other variables, such dynamics being observed over a timescale $\tau$ (maybe tunable).
As an example, the simplest way to define a causal network is the correlation matrix, however not always the most appropriate. 

To quantify such degree of causation we motivate the use of our information response, defined as the ratio of the change of a prediction over the change of a predictor, both evaluated as KL divergences. It has the form of an information-theoretic fluctuation-response theorem, and therefore it has both the invariance properties from information theory and the physical interpretation of a propagation of perturbations.
While the present setting is limited to dyadic relations between variables, a generalization in terms of simultaneous perturbations of multiple variables is possible, and will be discussed in a future manuscript. 

Once a particular definition of causation is chosen, determining and quantifying the strength of causal links becomes a problem of statistical estimation, and is the subject of causal inference \cite{runge2018causal,spirtes2016causal}. In this manuscript we are interested in the former problem, i.e., to \text{define} a quantitative measure of causation.




\end{widetext}
\bibliography{Supplementary}